\documentclass[a4paper,fleqn,usenatbib]{mnras}

\usepackage{newtxtext,newtxmath}

\usepackage[T1]{fontenc}
\usepackage{ae,aecompl}

\usepackage{graphicx}	
\usepackage{amsmath}	
\usepackage{amssymb}	
\usepackage{lscape}
\usepackage{rotating}
\usepackage{float}
\hypersetup{colorlinks = true, linkcolor = blue, urlcolor=blue, citecolor=blue} 
\usepackage[normalem]{ulem}
\usepackage{tabularx}
\usepackage{rotating, booktabs}
\usepackage{inputenc}
\usepackage{geometry}
\usepackage{changepage}
\usepackage{color}

\def\msun{\hbox{M$_\odot$}}
\def\bi{$m_{\rm F438W}-m_{\rm F814W} \,$}
\def\i{$m_{\rm F814W} \, $}
\def\b{$m_{\rm F438W} \, $}
\def\dFe{$\Delta$[Fe/H]}

\definecolor{mre}{rgb}{0.0,0.6,0.0}

\title[Fornax 4]{Is Fornax 4 the nuclear star cluster of the Fornax dwarf spheroidal galaxy?}

\author[Martocchia et al.]{S. Martocchia$^{1,2}$, E. Dalessandro$^{3}$, M. Salaris$^{2}$, S. Larsen$^{4}$, M. Rejkuba$^{1}$.
\\
$^{1}$European Southern Observatory, Karl-Schwarzschild-Stra\ss e 2, D-85748 Garching bei M\"unchen, Germany\\
$^{2}$Astrophysics Research Institute, Liverpool John Moores University, 146 Brownlow Hill, Liverpool L3 5RF, UK\\
$^{3}$INAF-Osservatorio di Astrofisica \& Scienza dello Spazio, via Gobetti 93/3, I-40129, Bologna, Italy\\
$^{4}$Department of Astrophysics/IMAPP, Radboud University, P.O. Box 9010, 6500 GL Nijmegen, The Netherlands\\}

\date{Accepted XXX. Received YYY; in original form ZZZ}

\pubyear{2020}

\begin{document}
\label{firstpage}
\pagerange{\pageref{firstpage}--\pageref{lastpage}}
\maketitle

\begin{abstract}
Fornax 4 is the most distinctive globular cluster in the Fornax dwarf spheroidal. Located close to the centre of the galaxy, more metal-rich and potentially younger than its four companions (namely, Fornax clusters number 1, 2, 3 and 5), it has been suggested to have experienced a different formation than the other clusters in the galaxy. 

Here we use HST/WFC3 photometry to characterize the stellar population content of this system and shed new light on its nature. 
By means of a detailed comparison of synthetic horizontal branch and red giant branch with the observed colour-magnitude diagrams, we find that this system likely hosts stellar sub-populations characterized by a significant iron spread up to \dFe$\sim$0.4 dex and possibly by also some degree of He abundance variations $\Delta$Y$\sim0.03$. 
We argue that this purely observational evidence, combined with the other peculiarities characterizing this system, supports the possibility that Fornax 4 is the nuclear star cluster of the Fornax dwarf spheroidal galaxy. A spectroscopic follow-up for a large number of resolved member stars is needed to confirm this interesting result and to study in detail the formation and early evolution of this system and more in general the process of galaxy nucleation.    

\end{abstract}

\begin{keywords}
galaxies: star clusters: individual: Fornax 4 $-$ galaxies: individual: Fornax dwarf spheroidal $-$ Hertzprung-Russell and colour-magnitude diagrams $-$ stars: abundances $-$ technique: photometry
\end{keywords}



\section{Introduction}
\label{sec:intro}

Fornax and Sagittarius are the only known dwarf spheroidals (dSphs) in the Local Group
that host a globular cluster (GC) system. Interestingly, despite having a relatively small stellar mass of a few times $10^7$\msun \citep{mcconnachie12}, the Fornax dSph hosts at least five GCs, with a sixth candidate recently confirmed as a likely cluster \citep{wang19}.

The Fornax GC system is interesting in many aspects: its GC specific frequency, i.e. the number of clusters normalised to the total visual magnitude of the galaxy, is among the largest observed \citep{georgiev10}. Even more intriguing is the observed difference in the peak metallicity of its GCs compared to the underlying stellar metallicity \citep{larsen12a}, the former being more metal-poor ([Fe/H]$\simeq -$2 dex) than the field stars ([Fe/H]$\simeq -$1 dex). These properties of the Fornax GC system provide important constraints for GC formation efficiency as well as mass-loss and self-enrichment \citep{larsen12a,lamers17}. 
 
Similarly to what is observed in ancient GCs in the Milky Way (MW), the Fornax clusters show light-element chemical abundance variations 
among
their stars, which are referred to as multiple stellar populations (MPs, e.g. \citealt{gratton12} and more references below).
MPs in GCs show specific patterns in their light elements, such as the Na-O, C-N, and (sometimes) Mg-Al anticorrelations (e.g. \citealt{marino08, carretta09a,cannon98,carretta09b}). 
MPs have been discovered in every environment studied to date, from the MW (e.g. \citealt{piotto15, gratton12, nardiello18}), to the Magellanic Clouds (MCs, e.g. \citealt{mucciarelli09,dalessandro16,niederhofer17a,gilligan19}), Sagittarius dwarf galaxy (e.g. \citealt{carretta10}), and M31 (e.g. \citealt{schiavon13}). Additionally, chemical anomalies have been studied in many clusters of different masses and ages, showing that both parameters may play a role. Indeed, recent works found that MPs are not only restricted to the ancient ($>10$ Gyr) GCs but have also been found in young massive star clusters down to $\sim$2 Gyr (in the MCs, \citealt{niederhofer17b,martocchia17,martocchia18a,martocchia19,hollyhead18,hollyhead19}), with the older clusters having larger N abundance spreads compared to the younger ones (at constant cluster mass, \citealt{martocchia19}). Similarly, for the ancient GCs, a correlation between abundance spreads and the mass of the cluster are observed (e.g., \citealt{schiavon13,milone17}), with the most massive clusters showing larger abundance variations.

To constrain the presence of MPs in Fornax clusters, \cite{larsen14} analysed the width of the red giant branch (RGB) of Fornax 1, 2, 3 and 5 by using HST filters sensitive to N variations finding that all four clusters host MPs in the form of N spreads. Spectroscopically, \cite{letarte06} studied abundances of individual stars in Fornax 1, 2 and 3 and report a Na-O anticorrelation as well as a marked similarity with Galactic GCs.
Additionally, \cite{dantona13} 
examined the morphology of the horizontal branch (HB) of the same GCs concluding that such systems must host a large fraction of He-rich stars. 

To date, no investigation 
of MPs in the cluster Fornax 4 has been performed. This cluster was likely excluded from previous studies due to the high contamination from field stars as in fact, this system is located very close to the galaxy center. The first and only hint that Fornax 4 might host chemical variations was given by \cite{larsen12}. They calculated the [Mg/Fe] from the integrated light spectra of Fornax 3, 4 and 5, finding it to be significantly lower than the [Ca/Fe] and [Ti/Fe] ratios, contrary to what is generally observed in field stars in the Galaxy and in dSphs. They interpreted this as a potential signature of MPs. 

By using optical Hubble Space Telescope (HST) photometry,  \cite{buonanno99} 
found that Fornax 4 has a much redder HB and brighter sub-giant branch (SGB) than the other GCs in Fornax. They 
concluded that Fornax 4 is $\sim$3 Gyr younger than the other clusters, which are on average $\sim$12 Gyr old. 
They derived a metallicity of [Fe/H]$<-2$ dex for Fornax~4, which is significantly lower than what obtained by \cite{strader03} based on integrated spectroscopy ([Fe/H]$=-$1.5 dex) and by \cite{larsen12} ([Fe/H]$=-$1.4 dex). On the contrary, the other GCs in the Fornax dSph are more metal-poor, with metallicities ranging from [Fe/H]$=-$1.8 dex to [Fe/H]$=-$2.3 dex \citep{larsen12}.
These results have been recently confirmed by \cite{deboer16} who studied the star formation history of Fornax 4 and found that it
is indeed younger and more metal-rich compared to the other clusters in the galaxy. 

The position, higher metallicity and younger age of Fornax 4 led many authors (e.g. \citealt{hardy02,strader03}) to consider it as the nuclear star cluster (NSC) of the Fornax dSph. A NSC is a very dense and massive star cluster which resides in the innermost region of a given galaxy \citep{boker02,neumayer11}. If this is the case for Fornax 4, it should be expected to show a significant iron spread, as it is typically observed in such systems (e.g. \citealt{walcher06,lyubenova13,kacharov18}).
However, at the moment there is no consensus about the real nature of Fornax 4 and whether or not it is a genuine GC or a NSC is still an open question (see the discussion in \citealt{hendricks16} for more details). 

By using HST/WFC3 archival observations, here we study in detail the stellar population properties of Fornax 4 with the aim of providing new clues on its nature and formation.
This paper is structured as follows: in Section \S \ref{sec:obs} we report on the photometric reduction procedures, while we outline the calculation of the structural parameters in Section \S \ref{sec:structure}. We estimate the age of Fornax 4 in Section \S \ref{sec:age}. In Section \S \ref{sec:mp} we characterize the stallar population properties in the system. Finally, we discuss and conclude in \S \ref{sec:disc}.

\section{Observations and data reduction}
\label{sec:obs}

We used HST/WFC3 images obtained trough filters F438W and F814W (GO-13435, P.I. M. Monelli).
The dataset consists of:  
i) 12 exposures of 200s each for the F438W, ii) 6 exposures of 150s each for the F814W.

The images have been processed, flat-field corrected, and bias-subtracted 
using the standard HST pipeline (flc images). Pixel-area effects have been corrected by using the Pixel Area Maps images. We also corrected all images for cosmic rays contamination by using the L.A. Cosmic algorithm by \cite{vandokkum01}.

The photometric analysis has been performed following the same steps as in \cite{dalessandro14}. We used DAOPHOTIV \citep{stetson87} independently on each chip and filter. We selected several hundred bright and isolated stars in order to model the point-spread function (PSF). The PSF was left free to vary spatially to the first-order. In each image, we then fit all the star-like sources with the obtained PSF as detected by using a threshold of 3$\sigma$ from the local background.
The final star lists for each image and chip have been cross-correlated by using DAOMATCH, then the magnitude mean and standard deviation measurements were obtained through DAOMASTER. We obtained the final catalog by matching the star lists for each filter by using DAOMATCH and DAOMASTER.
Instrumental magnitudes have been converted to the VEGAMAG photometric system by using the prescriptions and zero-points reported on the dedicated HST web-pages\footnote{see \url{http://www.stsci.edu/hst/wfc3/phot_zp_lbn}}. Instrumental coordinates were reported on the absolute image World Coordinate System by using
a catalogue centred on the centre of Fornax 4 (see \S \ref{sec:structure}) downloaded from the Gaia Archive\footnote{\url{https://gea.esac.esa.int/archive/}} and by means of the software CataXcorr\footnote{Part of a package of astronomical softwares (CataPack) developed by P. Montegriffo at INAF-OAS.}. 

\begin{figure}
\centering
\includegraphics[scale=0.4]{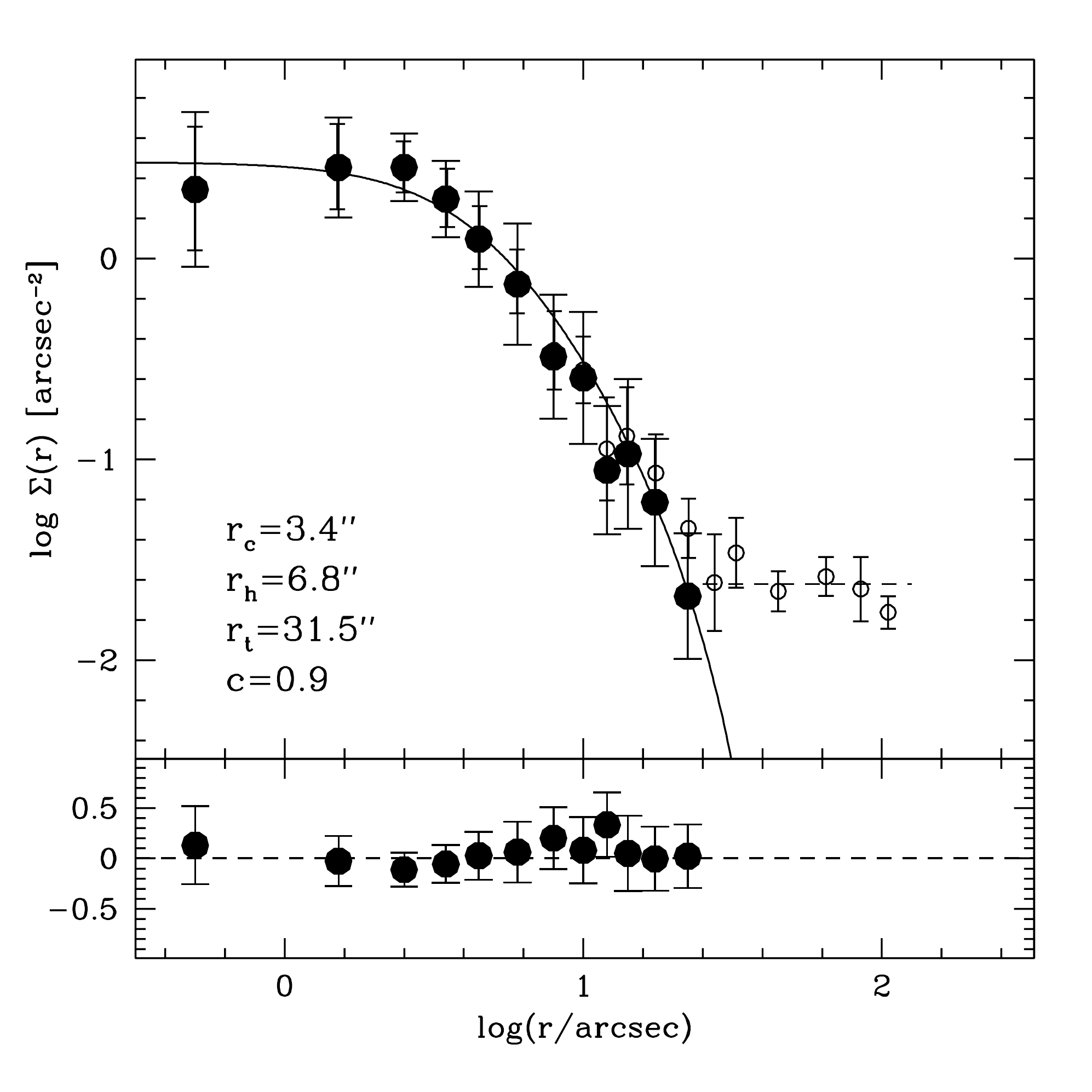}
\caption{\textbf{Top panel}: Stellar density of Fornax 4 as a function of radius. The open circles indicate the observed density profile while the black filled circles represent the background subtracted stellar density profile. The black solid line represents the best fit of the King profile. The dashed line indicates the estimation of the background. \textbf{Bottom panel}: the residuals of the fit are shown. See the text for more details.} 
\label{fig:radprof}
\end{figure}

\subsection{Artificial Stars Test}
\label{subsec:AS}

We performed artificial star (AS) experiments following the method described in \citeauthor{dalessandro11} (\citeyear{dalessandro11}, see also \citealt{bellazzini02,dalessandro15,dalessandro16}) to derive a reliable estimate of the photometric errors. 
Briefly, we generated a catalog of simulated stars with a F814W-band input magnitude (F814W$_{in}$) extracted from a luminosity function (LF) modeled to reproduce the observed LF in that band and extrapolated beyond the observed limiting magnitude. We then assigned a  
F438W$_{in}$ magnitude to each star extracted form the luminosity function, by means of an interpolation along the mean ridge line obtained from the observed $m_{F438W}-m_{F814W}$ vs $m_{F814W}$ colour magnitude diagram (CMD).
Artificial stars were added to real images by using the software  DAOPHOTIV/ADDSTAR. We minimized ``artificial crowding'', placing stars into the images following a regular grid composed by $15\times 15$ pixel cells (roughly corresponding to 10 FWHM) in which only one artificial star for each run was allowed to lie.
More than 100,000 stars have been simulated in each WFC3 chip. AS experiments had adopted the same reduction strategy and models for PSF that were used for real images on both real and simulated stars. In such a way, the effect of radial variation of crowding on both completeness and photometric errors is accounted for.
The AS catalog was then used to derive photometric errors for HB and RGB stars, which will be used in the following analysis (see \S \ref{sec:mp}). 
The analysis of the AS stars was carried out applying the same cuts in photometric quality indicators (sharpness -- $sharp$) that have been  applied in the data (see Section \ref{sec:structure}).

\begin{figure}
\centering
\includegraphics[scale=0.42]{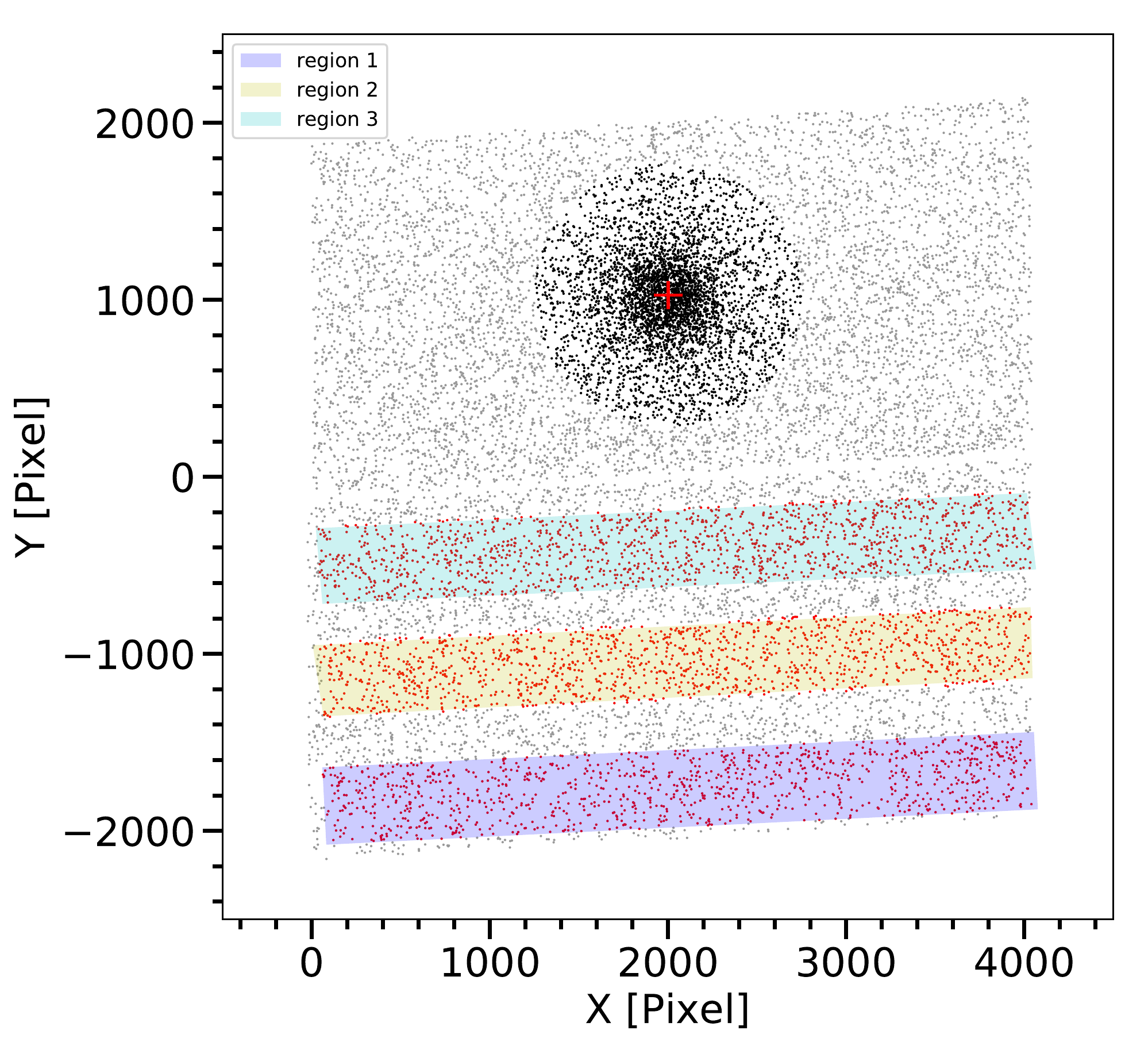}
\caption{Distortion-corrected X vs Y map for the WFC3 field. The red cross indicates the centre of the cluster while black circles represent stars that are within a radius of 30 arcsec from the centre of the cluster. The coloured regions represent the areas explored for the background, where stars used for the decontamination are indicated as red circles. Region 1 was finally addopted for the decontamination process. See text for more details.} 
\label{fig:xy}
\end{figure}

\begin{figure*}
\centering
\includegraphics[scale=0.4]{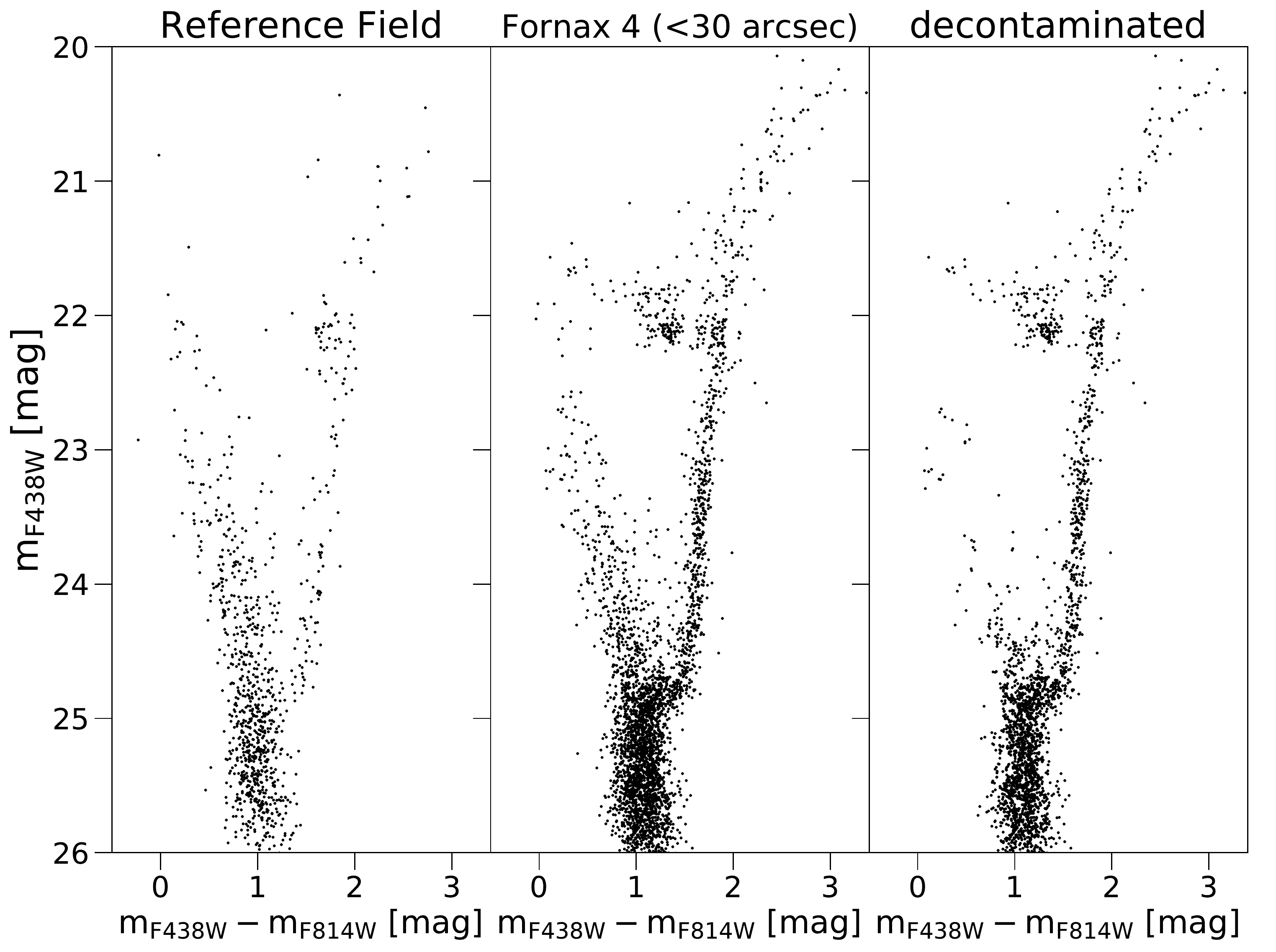}
\caption{Left panel: \bi vs. \b CMD of the reference field used for the decontamination. \bi vs. \b CMD of Fornax 4 before (central panel) and after (right panel) the field star subtraction.} 
\label{fig:fss}
\end{figure*}

\section{Structural Parameters}
\label{sec:structure}

To derive the structural parameters of Fornax 4, we built the cluster number density profile by using stars with sharpness $|sharp|<$0.1.
As a first step, we derived the centre of gravity of the cluster by using the same approach described in \cite{dalessandro13b}.
A first estimate of the cluster center
was performed by eye, then the center was 
measured through
an iterative procedure that averages the absolute positions of the stars lying within four different concentric radial regions ranging from 10$\arcsec$ to 25$\arcsec$ with a step of 5$\arcsec$. Only stars with with $m_{F438W}<25$ mag were selected.
The adopted cluster center is the mean of the different derived values, $C_{grav}$=(02:40:07.737,$-$34:32:10.96), with uncertainties $\sigma_{R.A.}=0.4\arcsec$ and $\sigma_{Dec}=0.3\arcsec$.
The density profile analysis was performed following the procedure fully described in \cite{miocchi13}. 

We used the AS catalogue to calculate the photometric completeness as a function of the distance from the cluster centre and magnitude. We assigned a completeness value $C$ to every star in the real catalogue.
We split the WFC3 FoV in 19 concentric annuli centered on $C_{grav}$, each one divided into two, three, or four sub-sectors. 
In each sub-sector, we estimated the total number of stars with $m_{F438W}<24$ normalized to their completeness, i.e. $\Sigma(1/C)$. 
The projected stellar density in each annulus 
is then the mean of the values measured in each sub-sector and the uncertainty has been estimated from the variance among the sub-sectors.  
The derived density profile is shown in Figure~\ref{fig:radprof} as open circles.  
The dashed line indicates the background, which was determined from stars at ${\rm Log}(r/{\rm arcsec})>1.4$. The black filled circles represent the background subtracted stellar density profile. We then derived the cluster structural parameters 
by fitting the observed density profile with a spherical, isotropic, single-mass \cite{king66} model. The best-fit model 
results in a cluster with a King dimensionless potential $W_0 = 5.0$, corresponding to a concentration parameter of $c=0.9,$
a core radius of $r_c = (3.4\pm0.7)\arcsec$ and 
a tidal radius $r_t=(31.5\pm4.8)\arcsec$. 

The structural parameters of the Fornax clusters were previously studied by \cite{webbink85} and more recently by \cite{mackeygilmore03}.
The latter determined the surface brightness profile of Fornax 4 by using WFPC2 observations in F555W and F814W bands. They find a core radius of (2.64$\pm0.27)\arcsec$ which is compatible within the errors with the value obtained in this work, although slightly lower.

Based on the obtained structural parameters, we then selected stars within a radius of $30\arcsec$ from the centre of Fornax 4, as ``cluster region''. Figure \ref{fig:xy} shows the instrumental coordinates (X,Y) map for Fornax 4 in the WFC3 field. Black points represent the selected stars in the cluster region while the red cross indicates $C_{grav}$. We performed a statistical decontamination analysis to get a clean CMD, following the method by \cite{niederhofer17b}. We defined a background reference region with the same area as the cluster region in order to statistically subtract field stars from the cluster CMD in the \bi vs. \b space. 
For every star in the background region, the closest star in colour-magnitude space in the cluster region is removed. 
Since the contamination in the field of Fornax 4 is large, we performed the field stars subtraction by using 3 different areas for the background region: one at the bottom of chip 2, one at the centre of chip 2 and a final one on the top. These regions are shown in Fig. \ref{fig:xy} with different colours. No significant differences were detected in the decontamination, by carefully comparing the decontaminated CMDs and reference fields among the three cases. We also performed additional tests to quantify the difference among the three regions. We selected RGB stars for each decontaminated catalogue, by using the same selection in magnitude and colour. We verticalised the RGB by using a fiducial line and calculated the distance of each star from the fiducial line to obtain a $\Delta$(colour)$\equiv\Delta$(\bi). The width of the RGB is estimated from the standard deviation of the $\Delta$(colour) (see \S \ref{subsec:metal}). The three different widths are the same, consistent up to the fourth digit. Thus our results are not affected by the choice of the decontamination region.  Additionally, we calculated the fraction of removed stars in each region and we found that this oscillates between 11 and 16\% (with a Poisson error of $\sim$8\%), demonstrating that there is no significant difference among them.
Hence, for the following analysis we decided to use the catalogue where the cluster and background regions are defined as shown in Fig. \ref{fig:xy}, with region 1 adopted for the background, because it is furthest away from the cluster and therefore least likely to contain any distant cluster members. Fig. \ref{fig:fss} shows the \bi vs. \b CMD of Fornax 4 (within 30$\arcsec$ from the cluster centre), before (central panel) and after (right panel) the field star subtraction. The left panel shows instead the CMD of the adopted reference field (i.e. region 1). While statistical decontamination may be prone to non negligible uncertainties (e.g. \citealt{dalessandro19}), it is possible to note how both the main sequence and red clump (\bi$\sim$1.5 mag and \b$\sim$22 mag) of the young population of stars in the Fornax dSph disappear after the correction. Finally, we report an additional test in Sect. \S \ref{subsec:metal}, to strengthen the argument that the results reported in this paper are not affected by the field star decontamination.

\begin{figure}
\centering
\includegraphics[scale=0.35]{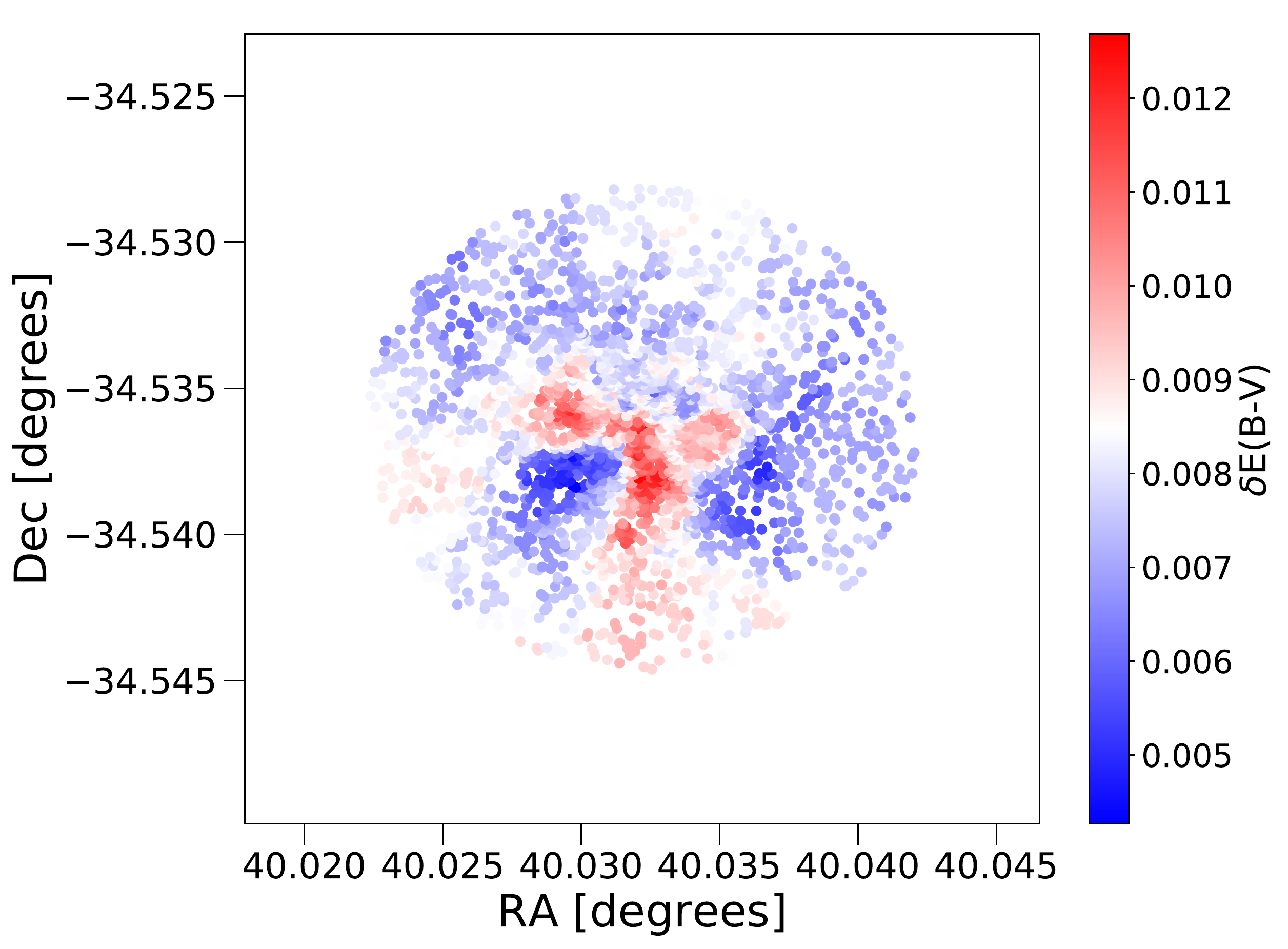}
\caption{Differential reddening map for Fornax 4 in the cluster region. See text for more details.} 
\label{fig:drmap}
\end{figure}

\subsection{Differential extinction}
\label{subsec:de}

\begin{figure}
\centering
\includegraphics[scale=0.35]{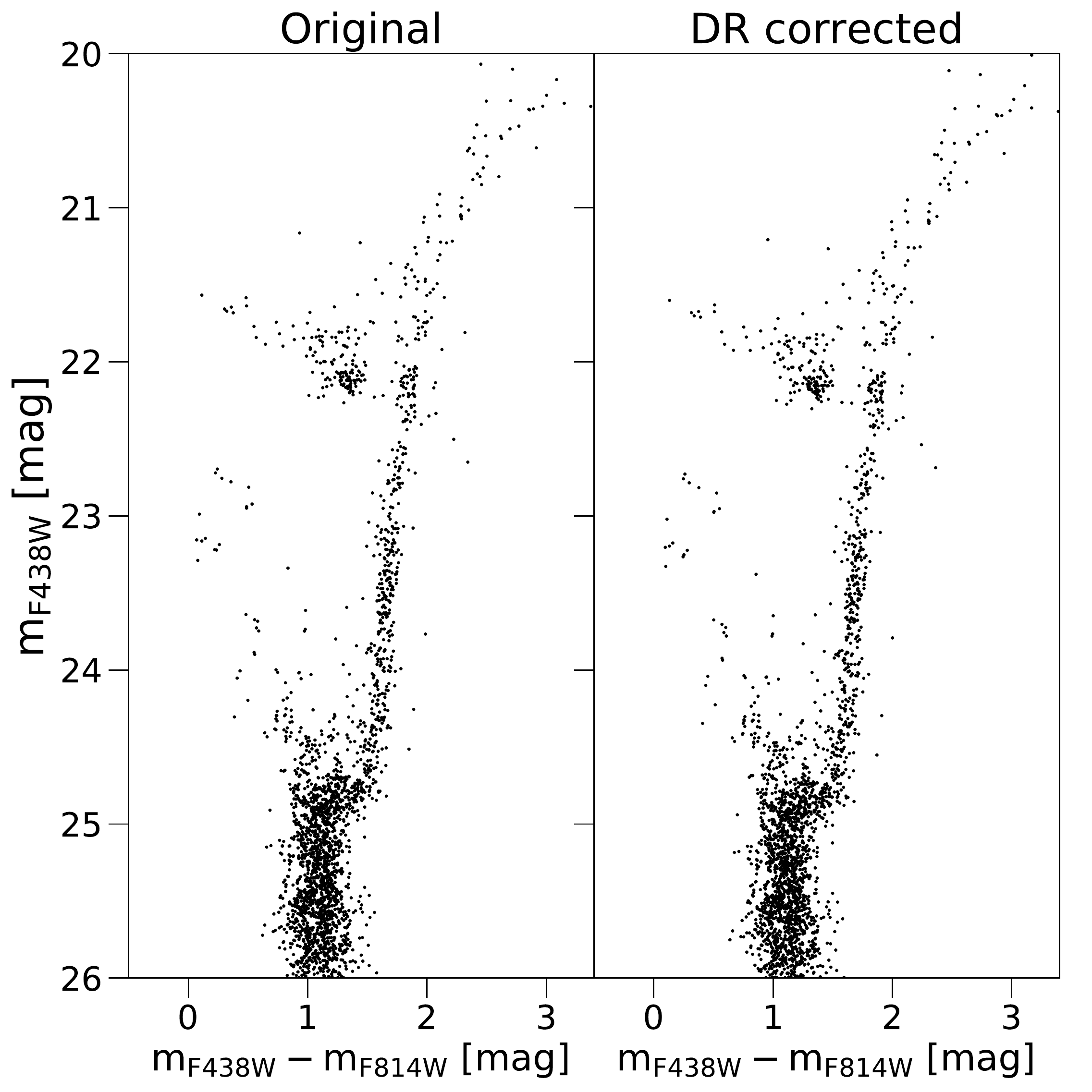}
\caption{Field stars subtracted \bi vs. \b CMDs of Fornax 4 before (left panel) and after (right panel) the differential reddening correction.} 
\label{fig:drcorr}
\end{figure}

We corrected our photometric catalogue for differential reddening (DR) by using the same method reported in \cite{dalessandro18} and \cite{saracino18}. We used our field stars subtracted catalogue for the estimation of the DR (see \S \ref{sec:structure}). 
We selected RGB stars in the magnitude range $22.5\lesssim$\b$\lesssim24.5$ and we defined a fiducial line in the \bi vs. \b CMD for these stars. 
We then calculated the geometric distance ($\Delta$X) from stars in this magnitude range that are 2$\sigma$ away from the line, where $\sigma$ represents the difference in colour between the stars and the fiducial line. 
For each star in the catalogue, the DR correction is then estimated by computing the mean of the
$\Delta$X values of the 20 nearest (in space) selected stars. By changing the number of neighbour stars (from 10 to 30), we obtain
very similar results.
The $\delta E(B-V)$ is obtained through the following equation:
\begin{equation}
\delta E(B-V) = \frac{\Delta X}{\sqrt{2R_{F438W}^2+R_{F814W}^2-2R_{F438W}R_{F814W}}},
\label{eq:dr}
\end{equation}
where $R_{F438W}=4.18$ and $R_{F814W}=1.86$ are the adopted extinction coefficients from \cite{milone15}. 
Fig. \ref{fig:drmap} shows the DR map for Fornax 4 in the cluster region (see \S \ref{sec:structure}), while Fig. \ref{fig:drcorr} shows the \bi vs. \b CMDs of Fornax 4 before (left panel) and after (right panel) the DR correction. We find a maximum $\delta E(B-V)$ of $\sim$0.013 mag, thus our catalogue is not significantly affected by differential extinction. Hereafter, we will use the DR corrected photometric catalogue. 

\begin{figure*}
\centering
\includegraphics[scale=0.45]{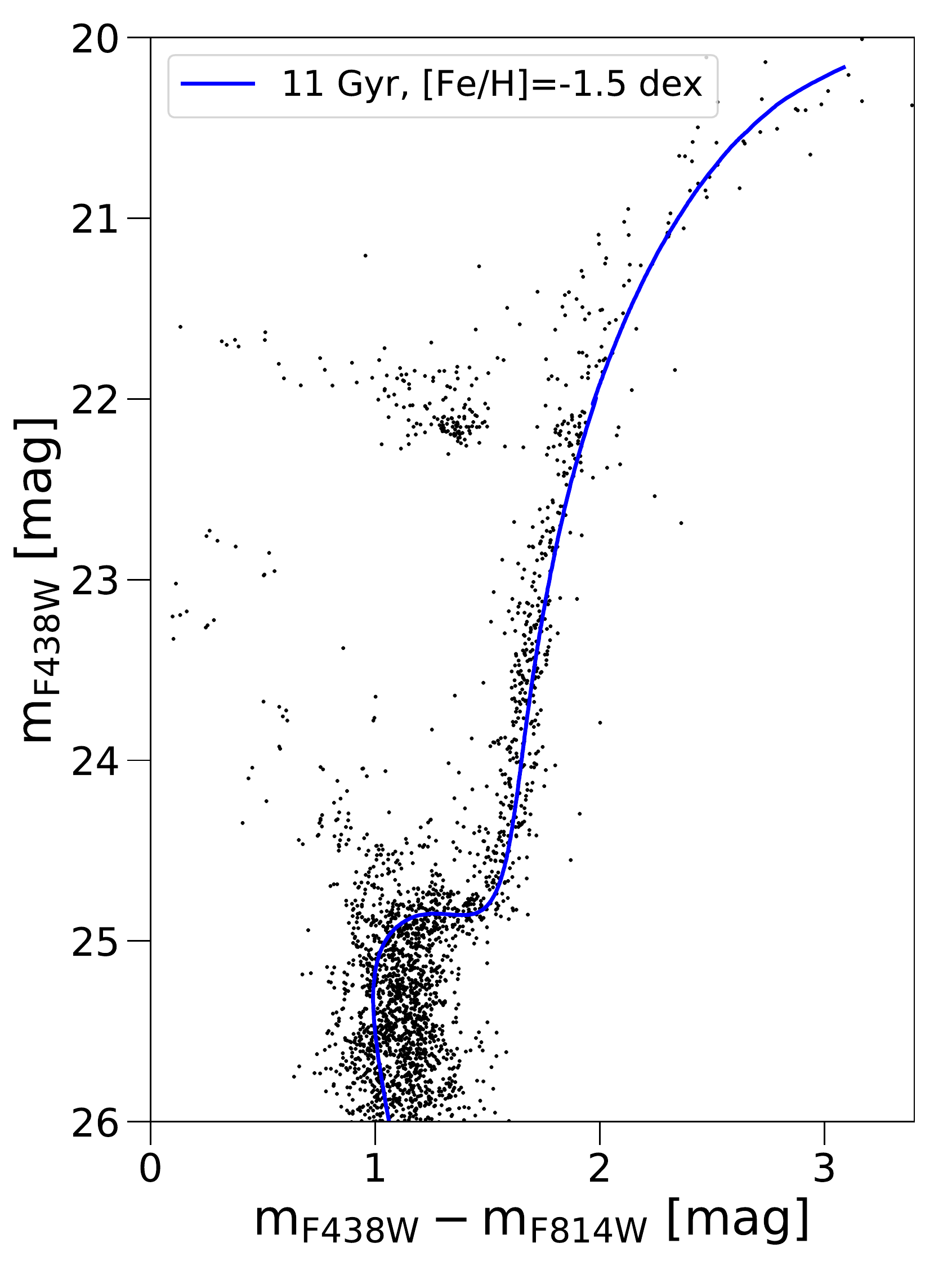}
\includegraphics[scale=0.45]{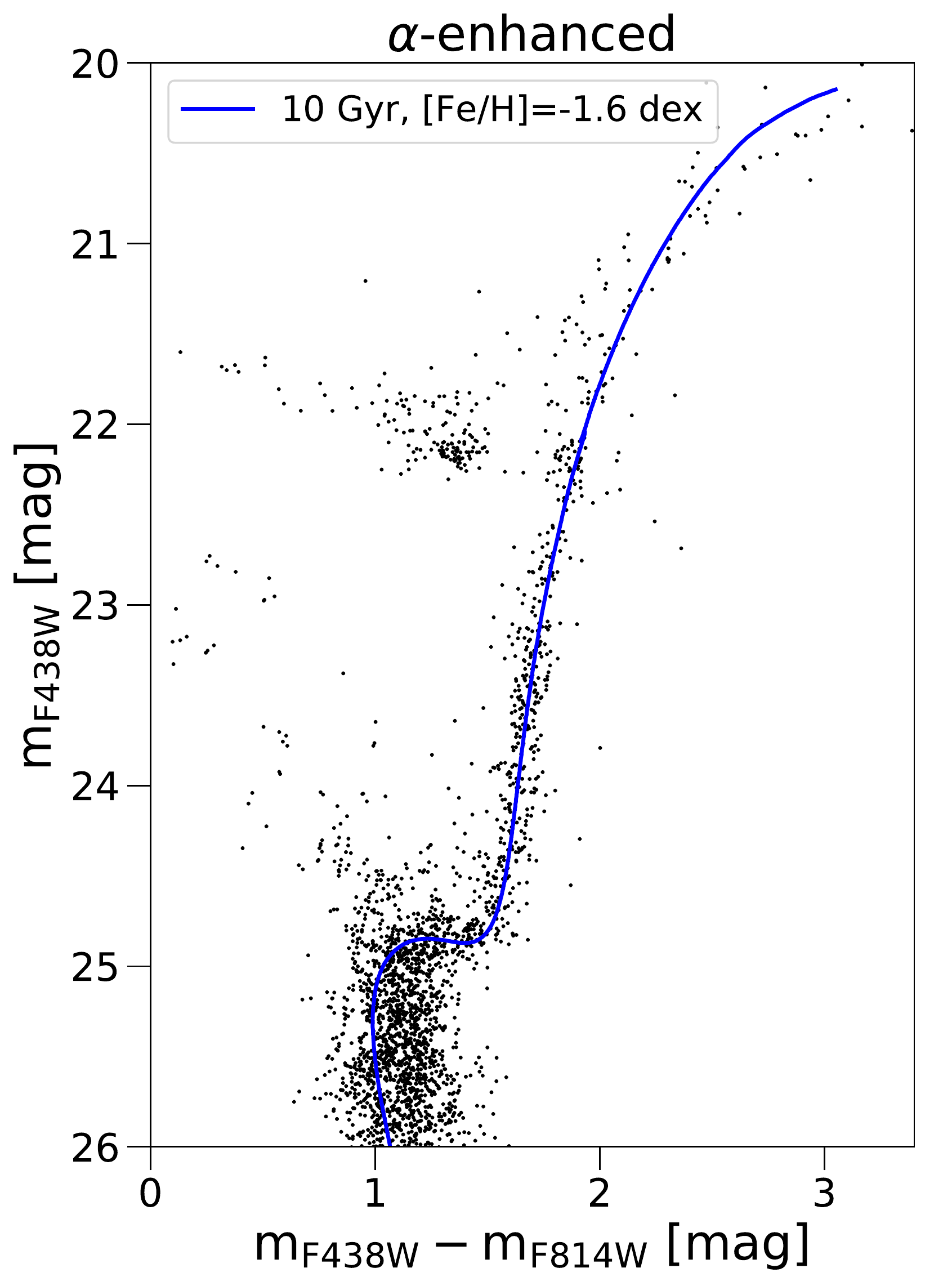}
\caption{\bi vs. \b CMDs for Fornax 4. The blue curve represents best fit solar (left panel) and $\alpha-$enhanced (right panel) BaSTI isochrone with the ages and metallicities shown in each panel. These are displayed with values of the extinction $E(B-V)=0.04$ mag and distance modulus $(m-M)_0=20.94$ mag.} 
\label{fig:age}
\end{figure*}

\section{Age and metallicity of Fornax 4}
\label{sec:age}

We used BaSTI isochrones (``A Bag of Stellar Tracks and Isochrones", \citealt{pietrinferni04}) in the \bi vs. \b CMD to 
obtain estimates
of the [Fe/H] and age of the cluster. For the absolute distance modulus we adopt the value ($m-M$)$_0=$20.94 mag obtained from the HB modelling (see discussion in \S \ref{subsec:hb}), while for the extinction we used $E(B-V)=$0.04 mag, which is in the range between 0.02 and 0.08 mag found in the literature (see \S \ref{sec:intro}). The extinction ratios employed to 
determine the extinction in the WFC3 filters have been calculated as described in \citet{gira08}, using the spectral energy distributions employed in BaSTI \citep{pietrinferni04}.

The best matching solar-scaled isochrone has [Fe/H]$=-1.5$ dex, and an age $t=11$ Gyr, whilst with $\alpha-$enhanced ([$\alpha$/Fe]=+0.4 dex, the only $\alpha$ enhancement available) isochrones we get [Fe/H]$=-1.6$  dex and $t=10$ Gyr (Fig. \ref{fig:age})\footnote{We note that the isochrone fit does not align perfectly for stars below the main sequence turnoff, however this is not affecting the results presented in this paper.}.
We are assuming that there is no chemical variation (in He and/or Fe) when estimating the age of the cluster.
In \S \ref{subsec:hb} we will discuss the presence of an initial He abundance and/or [Fe/H] spreads, using synthetic HB modelling and colour spread of the RGB. 
The derived distance modulus can change by a 
few 0.01~mag compared to $(m-M)_0$=20.94 mag, when these abundance spreads are included, but this does not affect substantially (less than 1~Gyr) the age estimates. 

The [Fe/H] values determined from Fig. \ref{fig:age} are in disagreement with \cite{buonanno99}, who find [Fe/H]$<-2$. 
However, our results agree well with the integrated light spectroscopy analyses by \cite{strader03} and \cite{larsen12}. Also, our solar-scaled age and metallicity are consistent with the work of \cite{hendricks16}. They used WFPC2 optical photometry and Dartmouth isochrones, finding a best fit of $t=$10 Gyr and [Fe/H]$=-1.5$ dex, assuming no $\alpha$-enhancement. 

Regarding the $\alpha$ elements, \cite{larsen12} report a small alpha-enhancement ([$\alpha$/Fe]$\sim+0.13$ dex) using integrated light spectroscopy, while \cite{hendricks16} report [$\alpha$/Fe]$=-0.19$ dex, although this result is based on a single member star, the only resolved star that has been studied so far in Fornax 4 spectroscopically. Given the current lack of consensus regarding the level of $\alpha$-enhancement present in Fornax 4 stars, we consider both isochrones in Figure \ref{fig:age} as best fits. 

Table \ref{tab:info} displays the information on Fornax 4 derived in this paper. 

\begin{table}
\caption{Properties of Fornax 4 derived in this work.}
\label{tab:info}
\centering          
    \begin{tabular}{c c}    
        \toprule\toprule
        Cluster  & Fornax 4 \\
        \midrule
        Age & 10$-$11 Gyr\\
        $[Fe/H]$ & -1.5 $-$ -1.6 dex \\
        $(m-M)_0$ & 20.94 mag\\
        $E(B-V)$ & 0.04 mag\\
        $r_c$ & 3.4 arcsec\\
        $r_t$ & 31.5 arcsec \\
       \bottomrule\bottomrule
        \end{tabular}
        \\
\end{table}

\section{Stellar population characterization}
\label{sec:mp}

In this Section we perform a detailed analysis of the RGB and HB population width and morphology to constrain the possible presence of sub-populations with different metallicity and/or He abundance.   

\subsection{The RGB width analysis}
\label{subsec:metal}

\begin{figure*}
\centering
\includegraphics[scale=1.0]{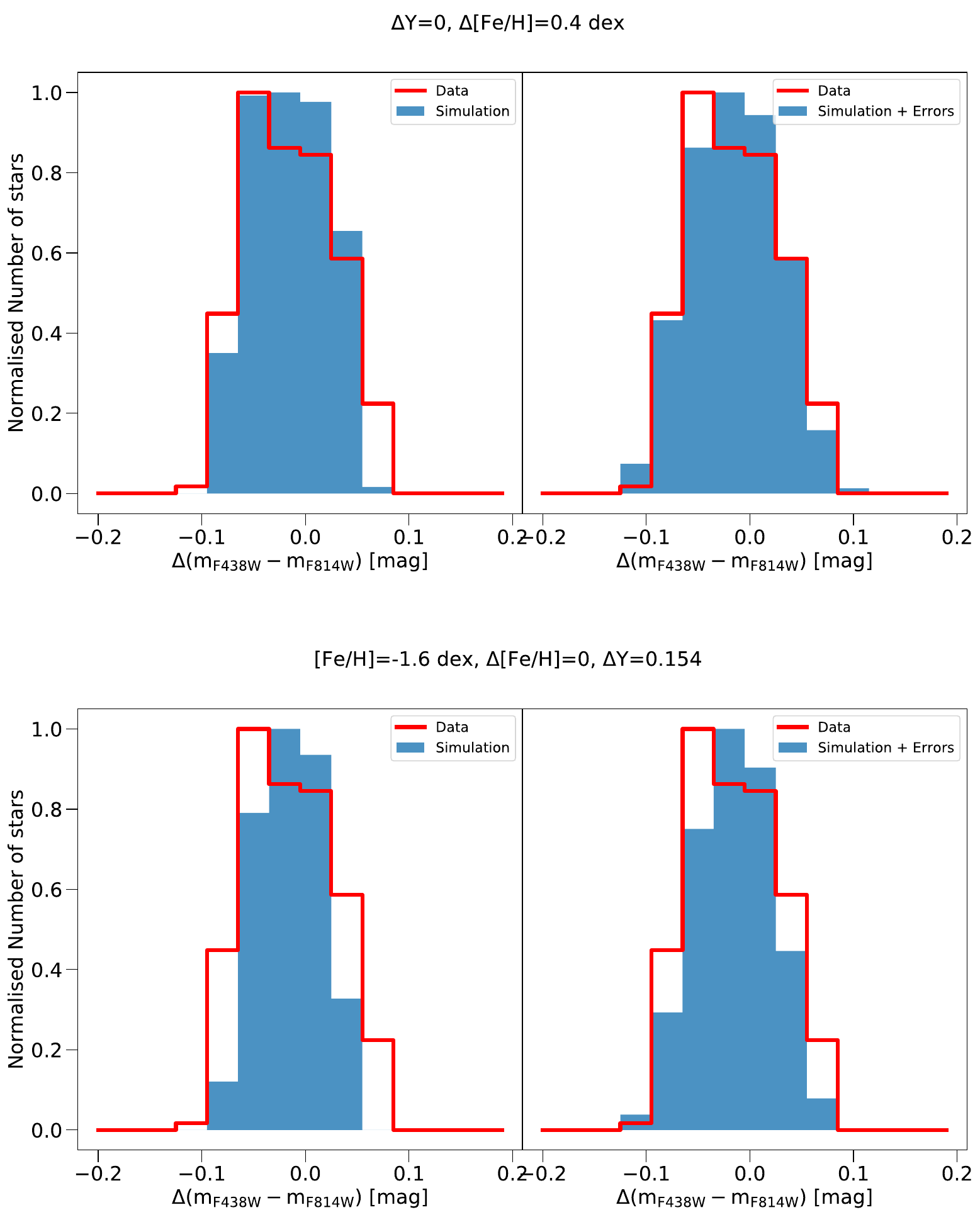}

\caption{Histograms of verticalised \bi colours vs. \b magnitudes for observed (red) and simulated (blue filled) RGB stars for $\Delta$Y=0, \dFe$=+0.4$ dex (top panels) and $\Delta$Y=0.154, \dFe$=0$ dex (bottom panels). See text for more details.} 
\label{fig:simu}
\end{figure*}

We focused our analysis on observed RGB stars in the \bi vs. \b CMD, in the magnitude range $22.6<$\b$<24.3$, i.e. the lower RGB, which is the more populated. We estimated an average error in this magnitude range from our AS tests (see \S \ref{subsec:AS}) and we obtained e(\b)$\simeq$0.018 mag and e(\i)$\simeq$0.012 mag\footnote{Errors were derived computing the r.m.s. of the distributions of simulated stars in the (mag$_{in}$, mag$_{in}$ $-$ mag$_{out}$) diagrams for the available bands in different magnitude bins (the RGB in this case) and after applying the same selections that were originally applied to the data. We calculated the distribution of the errors as a function of the distance from the cluster centre. To be conservative, the values we adopted for the errors are the maximum in each band, measured close to the cluster centre.}. 
We find that the observed RGB width in the selected magnitude range is $\sim0.042$ mag, hence it is significantly larger than what expected from photometric errors, i.e. e(\bi)$\simeq0.022$ mag. 

Since both He and Fe abundance variations affect stellar temperatures during the RGB phase, they are both expected to produce a broadening of the RGB.

First, we quantitatively estimated the value of \dFe\ needed to reproduce the width of the cluster RGB. 
We generated 500 isochrones with a uniform distribution in metallicity, by interpolating between the $\alpha$-enhanced BaSTI isochrone from [Fe/H]$=-1.6$ dex up to [Fe/H]$=-1.0$ dex (see Fig. \ref{fig:age} and Table \ref{tab:info}). We kept $\Delta$Y=0. We populated each isochrone of the distribution in such a way that the LF in F438W magnitudes of the observed RGB is reproduced.
We then added Gaussian noise to each isochrone according to the photometric uncertainties listed above, in order to simulate the 
RGB with a range of metallicities.
We let the spread \dFe\, varying. We compared the observed versus the simulated width of the RGB for spreads \dFe=+0.2, +0.3, +0.4 and +0.5 dex. We verticalised the observed and simulated RGBs by defining two different fiducial lines in the \bi vs. \b space. This is done to account for the different slope of the RGB between the observations and the theoretical isochrones. We then calculated the distance in \bi colours of each star from the respective fiducial line,  $\Delta$(\bi). 

The results are shown in the top panels of Fig. \ref{fig:simu}, where we
plot
the histogram of the distribution of the verticalised \bi colours vs. \b magnitudes for observed (red) and simulated (blue filled) RGB stars. The histograms are normalised to the maximum of the distributions. The left panel shows the comparison between the data and the simulations when no photometric errors are included. 
For each spread we performed a Kolmogorov-Smirnov (KS) test to compare the data and simulated distributions. We obtained the highest p-value ($\sim$ 65\%) when a spread \dFe=+0.4 dex is employed.

We repeated the same analysis on the RGB assuming now that there is no variation in Fe and investigating the possible presence of a He spread. Therefore we used isochrones at fixed metallicity ([Fe/H]$=-$1.6 dex) with different He content. We generated 500 isochrones with a uniform distribution in He, ranging from Y=0.246 to Y=0.4. We repeated exactly the same steps described above and we show the results in the bottom panels of Fig. \ref{fig:simu}. To reproduce the observed width of the RGB, at least a $\Delta$Y=0.154 is needed. While probably an even larger He variation would allow a better match with the observed RGB width, this is the maximum spread we can obtain with the available set of models.

It is important to stress that, 
when comparing observations to simulations that include errors estimated from the AS, such errors may be underestimated, thus the values we report for \dFe and $\Delta$Y are upper limits. The main reason is that all AS experiments are simplified to some extent and they are not able to account for all the instrumental sources of noise. The typical difference between errors from AS and true observational uncertainties has been estimated in previous studies and is of the order of $30-40$\% (see Fig. 4 of \citealt{dalessandro11} and related text and Fig. 21 of \citealt{milone12}). We repeated the same analysis above by using errors that are 30\% larger. According to the KS test, we still found
that the simulated distributions that best
reproduce the observations are the ones having \dFe=+0.4 dex ($\Delta$Y=0), and  $\Delta$Y=0.154 ($\Delta$[Fe/H]=0).

We can therefore safely conclude that either Fornax 4 hosts stars with significantly different metallicity, with a total iron abundance spread of \dFe=+0.4 dex, or stellar sub-populations with large He variations for a total $\Delta$Y=0.154.
A combination of sub-populations with smaller variations of Fe and He can also match the observed RGB colour distribution. For instance, we reproduced the width of the RGB by making a simulation that includes a spread in He $\Delta Y$=0.03 (which will be constrained in the next Section \S \ref{subsec:hb}) and a slightly lower iron spread \dFe$=$+0.3 dex.

Finally, in order to ensure that the broadening of the RGB is not caused by uncertainty in statistical decontamination by some residual field stars (see \S \ref{sec:structure}), we carried out the following test. We selected RGB stars within the same colour-magnitude limits for both the field and cluster region (left and right panels of Fig. \ref{fig:fss}, respectively). We used the same isochrone as in the right panel of Fig. \ref{fig:age}, i.e. representative of the cluster metallicity, to verticalise the RGB stars in both cluster and field. We then calculated the $\Delta$(colours) for both populations. We obtained that the distributions of cluster and field stars in colours, i.e. metallicity, are peaked at the same mean. This test shows that even if there were some residuals left from field stars in the RGB, the broadening that we measure cannot be entirely due to field stars because there is no significant displacement between the RGB of the cluster and the one representative of the field.

\begin{figure}
\centering
\includegraphics[scale=0.35]{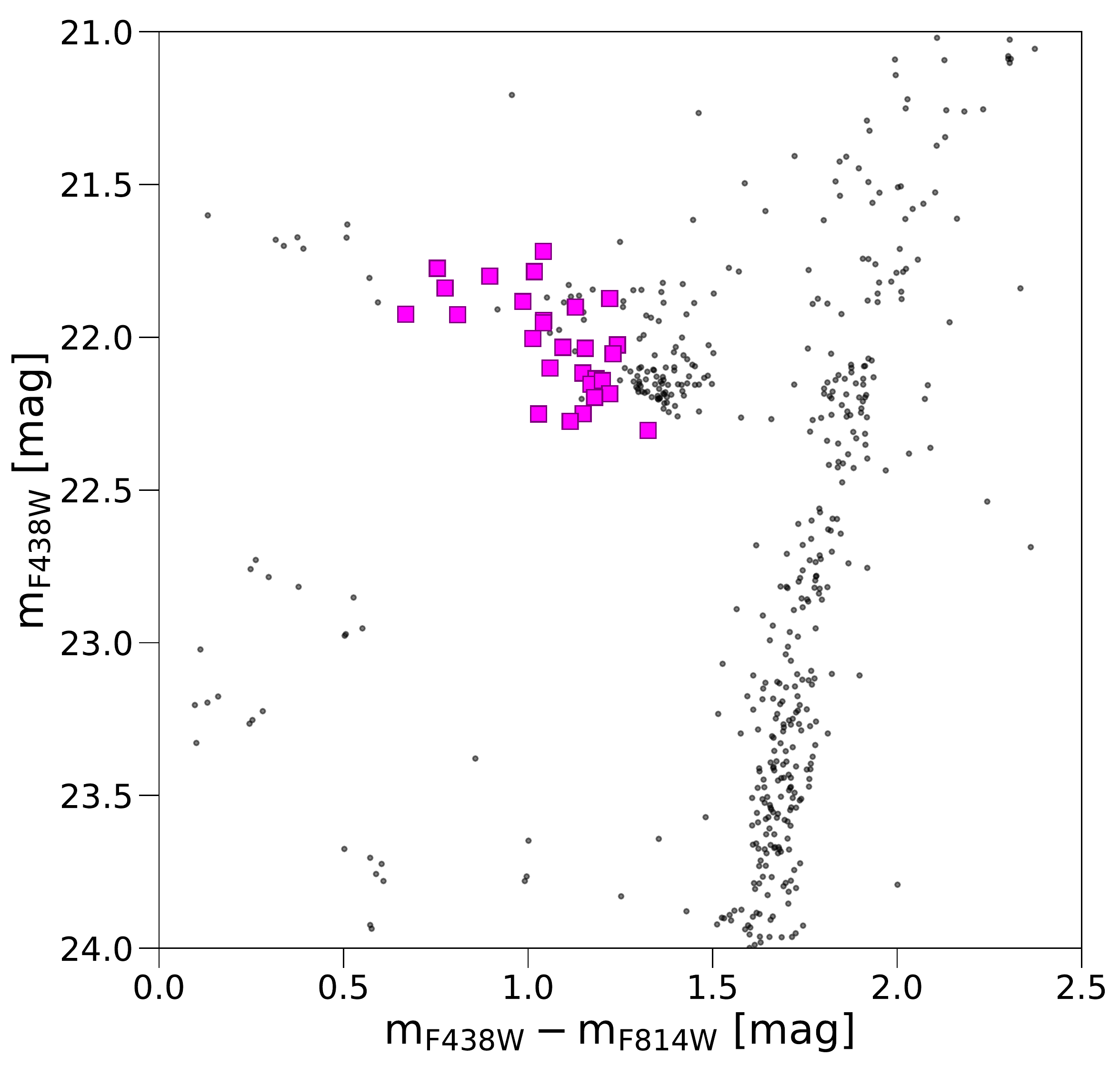}
\caption{Zoomed in \bi vs. \b CMDs for Fornax 4. Magenta squares represent variable stars identified in this work.} 
\label{fig:varstars}
\end{figure}

\subsection{The Horizontal Branch analysis}
\label{subsec:hb}
To try and constrain better the range of $Y$ and/or [Fe/H] spanned by the cluster initial chemical composition, we performed a detailed analysis of the horizontal branch, whose morphology is also affected by variations of the initial helium and metal content.
To this aim we used the same approach described in \cite{dalessandro11b,dalessandro13} which is based on the comparison between observations and synthetic HB models. 

Fornax 4 hosts a relatively large number of variable stars \citep{greco07}, for which we have only observations at random phase. Thus, 
before analysing the HB with stellar models, we needed to identify these stars in our catalogue, and remove them from the comparison. 


The first identification of variable stars in the Fornax clusters was performed by \cite{greco07}. By taking $B$ and $V$ time series photometry with MagIC on the Magellan Clay Telescope, they found 29 variable stars (out of which, 27 are identified as RR Lyrae), in a $2.4\arcmin\times2.4\arcmin$ area centred on Fornax 4. They claimed that the 22 stars located within the innermost 30$\arcsec$ are likely cluster members. 
Since we have several exposures in each filter in our dataset (\S \ref{sec:obs}), we used the variability index (VI) yielded by DAOPHOT to check for variable stars. 
We marked as ``variable" all stars having VI$>2$ both in the F438W and F814W band. Figure \ref{fig:varstars} shows a zoomed-in \bi vs. \b CMD around the HB region. Magenta squares represent the stars that are found to be variable in both bands simultaneously. In total we find 28 variable stars.

\begin{figure*}
\centering
\includegraphics[scale=0.4]{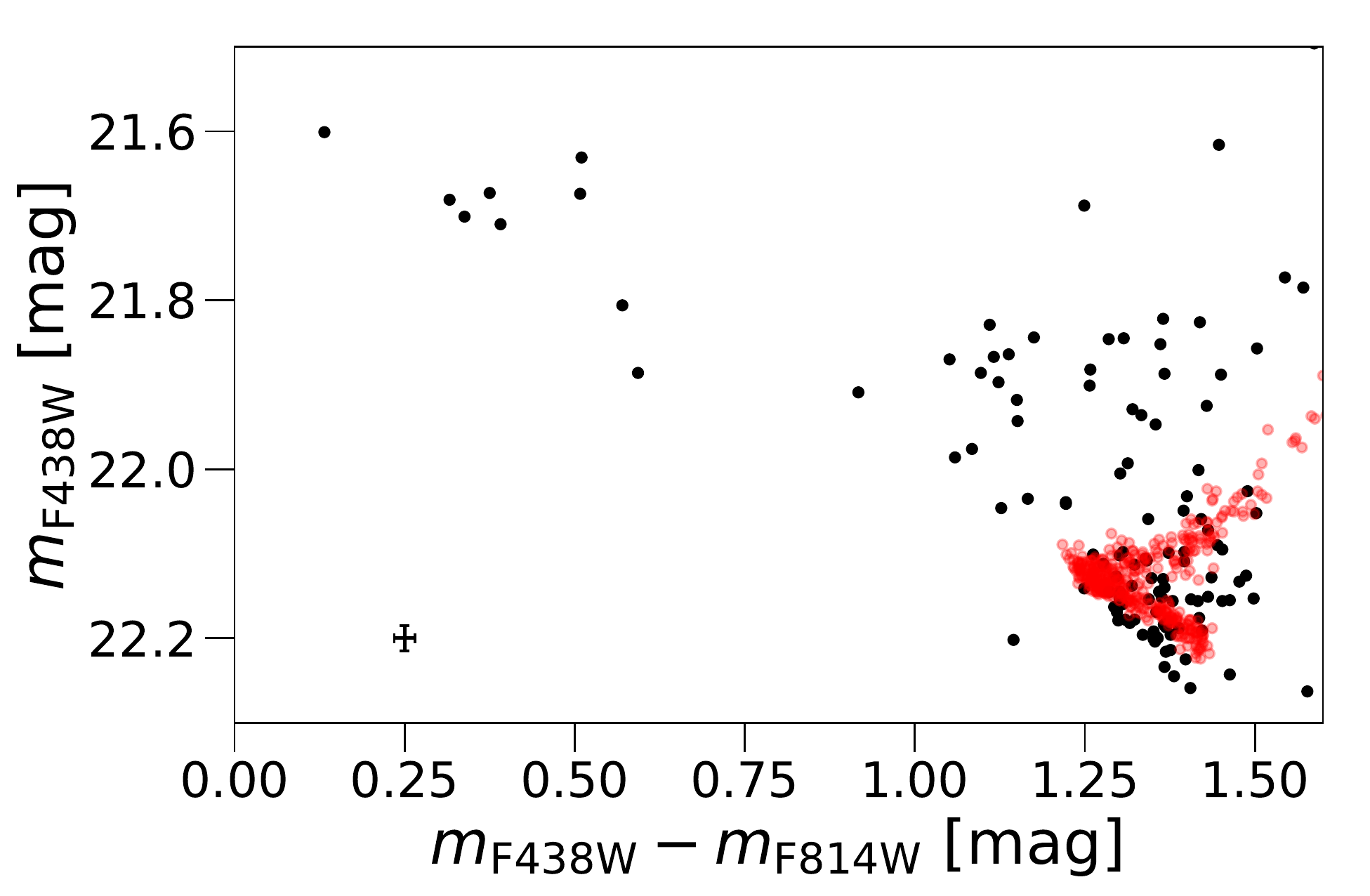}
\includegraphics[scale=0.4]{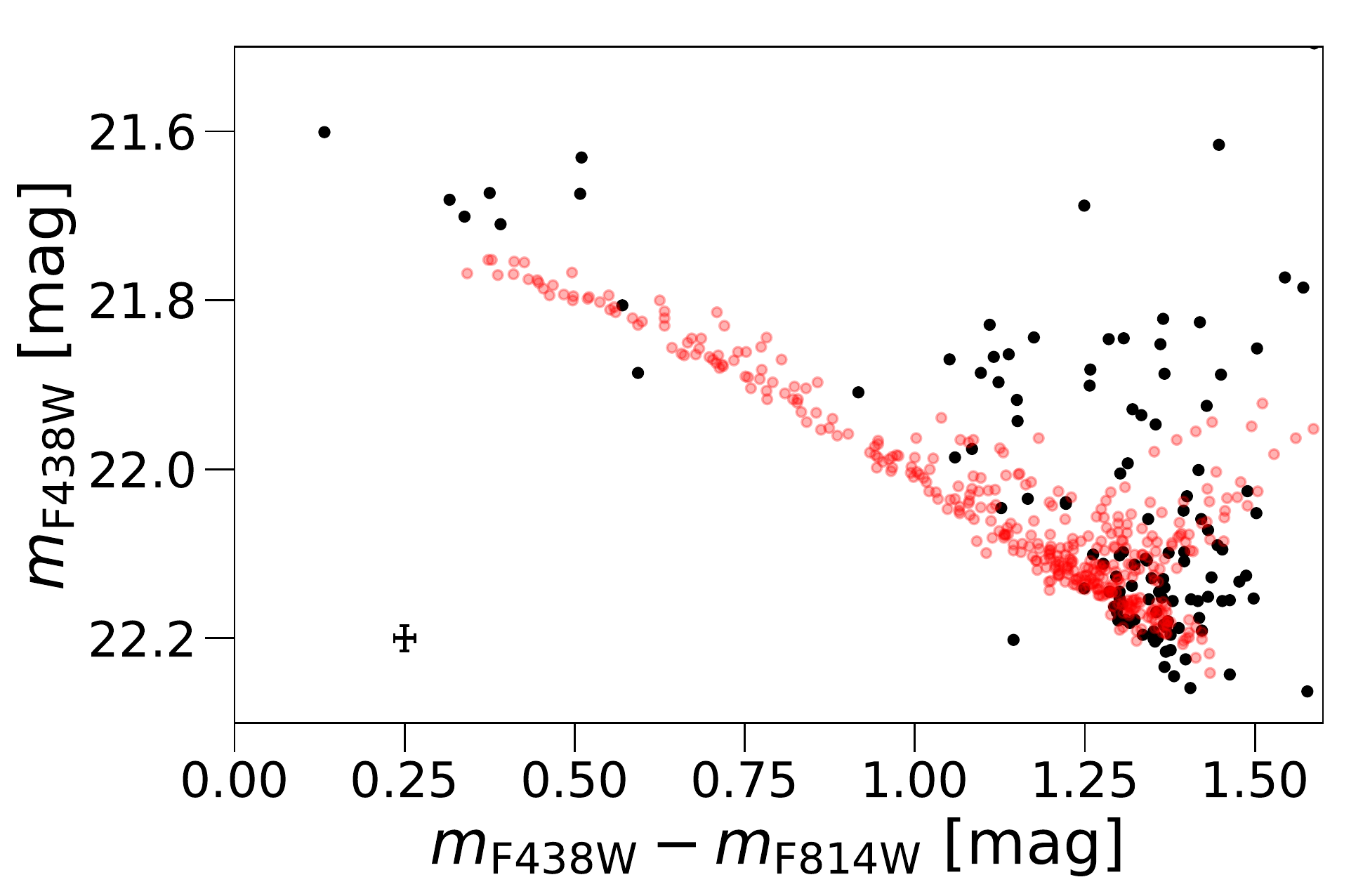}
\caption{\bi vs. \b CMD of HB stars in Fornax 4. Black circles denote observations while red circles represent synthetic HB stars. The left panel displays a synthetic HB 
calculated with $\Delta Y$=0, $\Delta M$=0.165\msun (with Gaussian $\sigma$=0.001 \msun  spread) and $(m-M)_0$=20.94 mag. The right panel shows a synthetic HB calculated with $\Delta Y$=0,  $\Delta M$ between 0.165 and 0.225 \msun (uniform distribution), $(m-M)_0$=20.94 mag. Average photometric errors are reported in the lower left corner. See text for more details.} 
\label{fig:massloss}
\end{figure*}

We matched our whole Fornax 4 photometric catalogue with the \cite{greco07} variable stars catalogue, in order to identify variable stars independently from our method (i.e. the VI index). We find 25 out of 29 stars in common. Of the remaining four, three stars are not in our WFC3 field of view while one star from the \cite{greco07} catalogue is not identified in the match. Out of these 25 stars in common, we were able to identify 19 stars as variables according to the VI index. 
We then removed all our 28 variable stars (pink squares in Fig. \ref{fig:varstars}) from the following analysis.

To assess the impact of He and Fe abundance spreads on the cluster HB we compared the observed \bi vs \b CMD with synthetic HB models. This technique  
has been already applied to several Galactic GCs (e.g., \citealt{dalessandro11b,dalessandro13}) and also Magellanic Clouds' clusters (e.g. \citealt{niederhofer17a}, \citealt{chantereau19}). 
For the synthetic HB calculations we used the BaSTI $\alpha$-enhanced HB models \citep{pietrinferni04, pietrinferni06} with metallicity [Fe/H]$=-1.6$ dex, and employed the code described in \citet{dalessandro13}. In our simulations with a [Fe/H] spread we have replaced the interpolation in $Y$ with an interpolation in [Fe/H], keeping the structure of the code unchanged. 
After assuming a reference age $t=$10 Gyr (that fixes the initial value of the mass 
currently evolving at the tip of the RGB), the only remaining parameters 
that determine the mass distribution (hence magnitudes and colours) along the synthetic HB are the total mass lost by the RGB progenitors $\Delta M$, the range of initial Y  ($\Delta$Y) or [Fe/H] ($\Delta {\rm [Fe/H]}$) values, and their statistical distribution. 
In our simulations we also input the 1$\sigma$ photometric errors as obtained from the AS test (see \S \ref{subsec:AS}). 
We notice that in terms of the mass distribution along the synthetic HB a  
variation of the cluster age can be compensated by changing $\Delta M$. For example, an age increase by 1 Gyr is compensated by a $\sim$0.02$M_{\odot}$ decrease of $\Delta M$.

As a first test, we checked whether a match of the observed HB morphology with theoretical models requires a spread of initial chemical composition.   
To this purpose, we have first calculated a synthetic HB with a small RGB mass loss,  
$\Delta M$=0.165\msun (and a Gaussian $\sigma$ spread equal to 0.001\msun). We assumed the same $E(B-V)$=0.04 mag employed in the isochrone fitting, and determined a cluster distance modulus by matching the peak of the number distribution of synthetic stars' magnitudes, to the observed one in the \bi colour range between 1.25 and 1.45~mag (the well populated red end of the observed HB distribution). In this way we have fixed the distance modulus also for the other simulations that follow. From the left panel of Fig. \ref{fig:massloss}, it is obvious that this simulation is not able to reproduce the full colour and magnitude extension of the observed HB. Hence, the right panel of Fig. \ref{fig:massloss} shows another synthetic HB, this time calculated with $\Delta M$ uniformly distributed between 0.165\msun and 0.225\msun, e.g. with a much larger mass loss spread. 
The colour extension is now well reproduced, but the synthetic HB is too faint to match the stars observed between \bi=0.25 and 0.50.
In these simulations and the ones that follow, the observed star count distribution as a function of colour is different from the synthetic ones. This is however not essential for our purposes, as we are not trying to perform a best fit of the HB.
This would be impossible given that 
the instability strip of the observed HB is depopulated, because we removed RR Lyrae variables for which we lack average magnitude measurements.
The goal of this analysis is to test whether the initial chemical composition scenarios inferred from the RGB are broadly consistent with the observed HB morphology.

\begin{figure}
\centering
\includegraphics[scale=0.4]{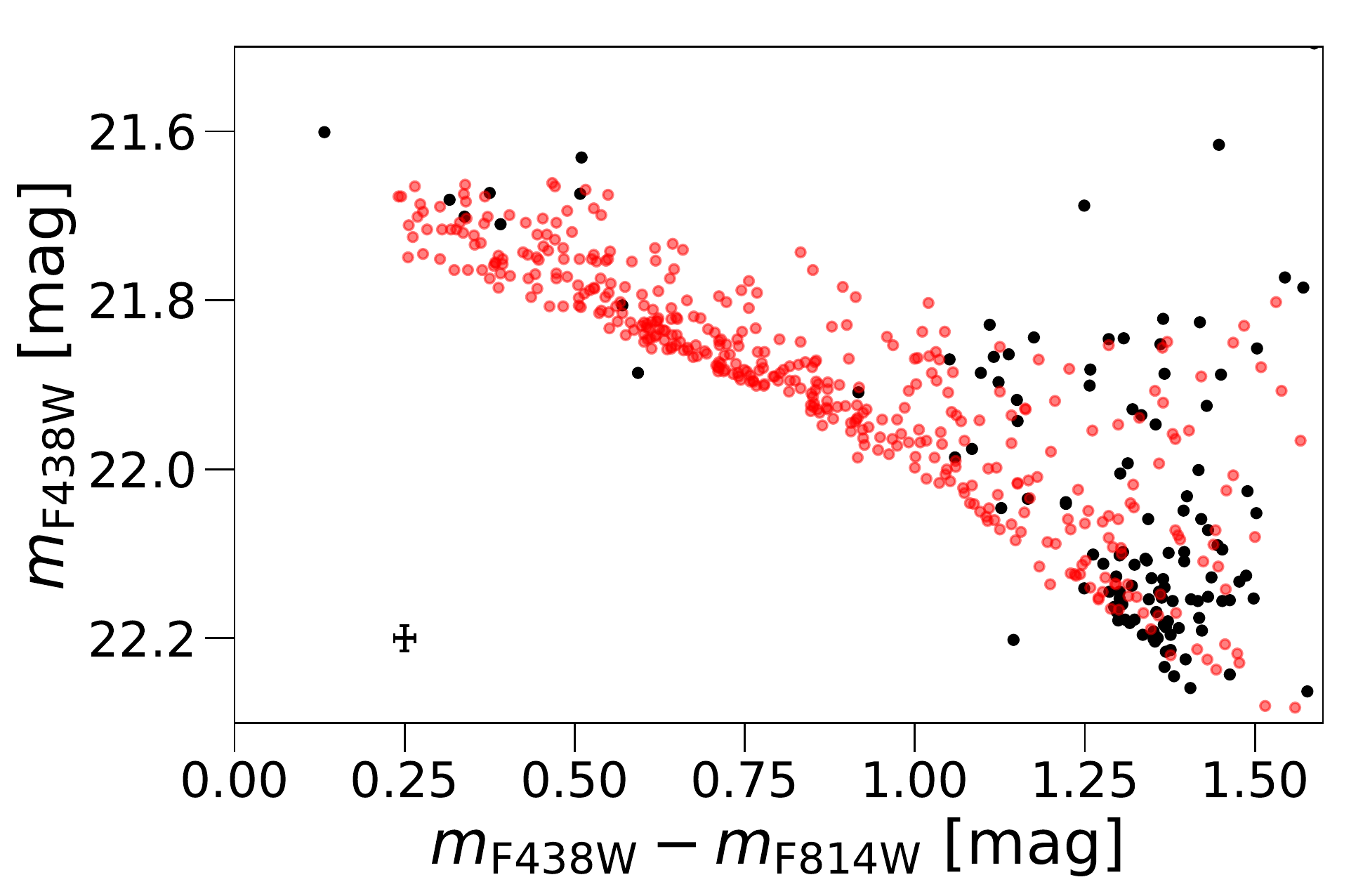}
\caption{As Fig.~\ref{fig:massloss}, but the synthetic model is calculated with a spread in iron $\Delta$[Fe/H]=+0.4 dex (uniform distribution), $\Delta Y$=0, $\Delta M$=0.23+0.06([Fe/H]+1.62)\msun, $(m-M)_0$=20.86 mag. See text for more details.} 
\label{fig:synthFe}
\end{figure}

As a second step we have examined whether the cluster HB can be reproduced by 
models with constant Y ($Y$=0.246) and \dFe$\sim+0.4$ dex, as derived from the RGB colour distribution. 
To this aim we have calculated a synthetic HB with a uniform probability [Fe/H] distribution between 
[Fe/H]=$-$1.62 dex and [Fe/H]=$-$1.22 dex, and a mass loss that increases linearly with [Fe/H] as $\Delta M$=0.23+0.06([Fe/H]=+1.62)\msun , and a 1$\sigma$ Gaussian dispersion of 0.005\msun\, around this mean relationship. This comparison is shown in Figure \ref{fig:synthFe}.
A constant mass loss irrespective of [Fe/H] produces a HB too extended in colour 
compared to the observations. Notice that in case of a [Fe/H] spread the metal poor component is located at the blue end of the synthetic HB.

\begin{figure}
\centering
\includegraphics[scale=0.4]{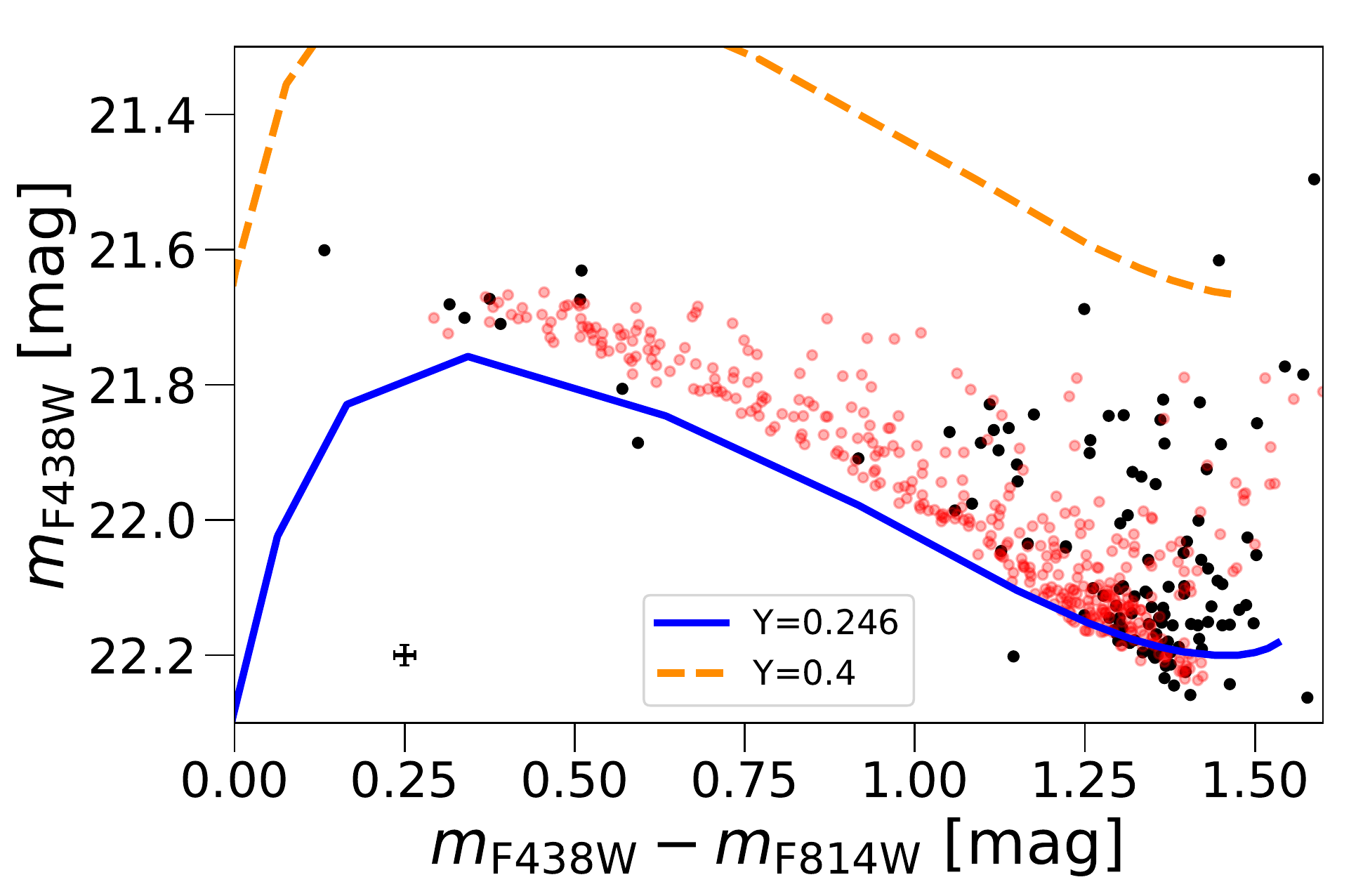}
\caption{As Fig.~\ref{fig:massloss}, but the synthetic HB stars are calculated with $\Delta Y$=0.03 (uniform distribution), $\Delta M$=0.160 \msun (Gaussian distribution with $\sigma$=0.003\msun), and $(m-M)_0$=20.98 mag. The solid blue and dashed orange curves represent the ZAHB for $Y=$0.246 and $Y=$0.4 at [Fe/H]$=-1.6$ dex, respectively. See text for more details.} 
\label{fig:HBhe}
\end{figure}

We have then checked whether models with a range of initial He abundances \
$\Delta Y$ (at constant [Fe/H]) compatible with the RGB constraint can also match the observed HB of Fornax 4.
Figure \ref{fig:HBhe} compares the observed HB with a synthetic one calculated including a He spread $\Delta Y$=0.03 (uniform probability distribution), 
and $\Delta M$=0.160\msun\, and Gaussian distribution with $\sigma$=0.003\msun.
The observed HB is overall well matched with this small value of $\Delta Y$, totally incompatible with 
the large $\Delta Y$ (at fixed [Fe/H]) inferred from the RGB. To make this point even clearer, the same figure shows for comparison also the zero age horizontal branch (ZAHB) for both $Y$=0.246 and $Y$=0.40. The $Y$=0.40 ZAHB is extremely overluminous compared to the data.

From this simple analysis, we are able to establish that a small spread of initial He abundances (up to $\Delta Y$=0.03) can also reproduce the shape of the HB of Fornax 4, but this spread is 
much lower than what derived from the RGB, assuming a constant [Fe/H].

The only way to achieve consistency between the RGB width and the HB morphology of Fornax 4 is to invoke either an initial spread of [Fe/H] (of about 0.4~dex) 
at constant $Y$, or both a small spread of $Y$ ($\Delta Y$ up to $\sim$0.03) 
and a spread of [Fe/H] of less than $\sim$0.4~dex. As an example, in Sect. \S \ref{subsec:metal} we have also reported that the width of the RGB can be reproduced by a combination of a spread in He $\Delta Y$=0.03 and slightly less iron spread \dFe$=$+0.3 dex.

\section{Discussion and conclusions}
\label{sec:disc}

In this work we investigated the nature of Fornax 4 by characterizing its stellar population properties. 
Indeed, because of its position, metallicity and age, this system has been suggested to be the nucleus
of the Fornax dSph galaxy (e.g. \citealt{hardy02,strader03}).

By using 
archival HST/WFC3 observations, we confirm that Fornax 4 is younger than the other clusters in the galaxy. In fact, we find the age of Fornax 4 is of the order of $t=11$ Gyr (or $t=10$ Gyr if $\alpha$-enhancement is present within the cluster).
We also find that Fornax 4 is more metal-rich than what previously found by  \cite{buonanno99} ([Fe/H]$=-$1.5/$-$1.6 dex) using optical CMDs, and in agreement with previous integrated light spectroscopic studies \citep{strader03,larsen12}.

We performed a detailed analysis of both the RGB and HB of Fornax 4 by means of a comparison between observations and synthetic CMDs. We find that the RGB and HB morphology can be simultaneously reproduced either by assuming  the presence of sub-populations 
with a spread in iron \dFe$\sim0.4$ dex or a combination of a slightly milder Fe spread and a variation of He abundance of $\Delta$Y$\sim$0.03 (see \S \ref{subsec:metal}, \S \ref{subsec:hb}).
While the exact amount of these variations may depend on model assumptions and the exact modeling of the photometric errors, this analysis clearly shows that {\it a non negligible iron spread is needed to reproduce the stellar population properties of Fornax 4}.  
This is a key information to assess the nature of this system. In fact, this result, in combination with its metallicity, position and age, provides support to the possibility that Fornax 4 is the NSC of the Fornax dSph. 

The most common scenarios invoke that NSCs form in-situ from the galaxy's central gas reservoir (e.g. \citealt{bekki07,antonini15,fahrion19}), or through GCs merging (e.g. \citealt{tremaine75,agarwal11,arcaseddacapuzzo14}), or through a combination of these (e.g. \citealt{hartmann11,antonini15,guillard16}). While the exact formation of NSCs is still debated (see \citealt{neumayer20} for a recent review), the general expected outcome is a system located at the center of the host galaxy which is characterized by the presence of sub-populations differing in terms of their iron abundances (e.g. \citealt{bekkifreeman03,bellazzini08}).
Additionally,  typical NSCs have a more extended star formation histories and some contribution from younger stars (e.g. \citealt{walcher05,kacharov18}). 
This seems not to be the case for Fornax 4, although it needs to be confirmed with a follow up study once the metallicity spread is fixed.
The star formation history of Fornax was recently derived by \cite{rusakov20} showing predominant intermediate age and old population ($5-10$~Gyr). If Fornax 4 sinked in the centre of the galaxy less than $\sim$5 Gyr ago, it would not have had much chance to accrete a substantial amount of gas and thus form additional stars.

Interestingly, it seems that Fornax 4 does not reside exactly in the kinematic centre of the galaxy, contrary to what it is found for M54, for instance. \cite{hendricks14} calculated the radial velocity (RV) of the Fornax dSph and this results to be $\sim$9 km/s higher than the RV of Fornax 4 (see \citealt{hendricks16} for a more detailed discussion). Nonetheless, if
the infalling of GCs is the dominant formation mechanism, finding a kinematic misalignment between the NSC and the center of the galaxy is expected (e.g. \citealt{capuzzomiocchi08,feldmeier14}).

While the interpretative scenario of Fornax 4 as the nucleus of the dwarf galaxy is extremely fascinating, it is necessary to confirm this result by performing a detailed spectroscopic and kinematic study of resolved member stars within the GC. It is important to note, in fact, that only one likely member star in the cluster has been analysed spectroscopically \citep{hendricks16} so far. This would provide a quantitative and reliable measurement of its stellar population chemical and kinematical patterns thus allowing a critical assessment of its formation and early evolution (see e.g. \citealt{sills19,alfaro19}), and more in general on the process on NSC formation and galaxy nucleation, being the second closest case after M54 \citep{bellazzini08}.

\section*{Acknowledgements}
We thank the referee for constructive comments that helped strengthen the paper.
S.M. is grateful for the warm hospitality of INAF-OAS Bologna where part of this work was performed and gratefully acknowledges financial support from the European Research Council (ERC-CoG-646928, Multi-Pop). E.D. aknowledges financial support by the project Light-on-Dark granted by MIUR through PRIN2017-000000 contract.  




\bibliographystyle{mnras}
\bibliography{f4} 

\begin{thebibliography}{}
\makeatletter
\relax
\def\mn@urlcharsother{\let\do\@makeother \do\$\do\&\do\#\do\^\do\_\do\%\do\~}
\def\mn@doi{\begingroup\mn@urlcharsother \@ifnextchar [ {\mn@doi@}
  {\mn@doi@[]}}
\def\mn@doi@[#1]#2{\def\@tempa{#1}\ifx\@tempa\@empty \href
  {http://dx.doi.org/#2} {doi:#2}\else \href {http://dx.doi.org/#2} {#1}\fi
  \endgroup}
\def\mn@eprint#1#2{\mn@eprint@#1:#2::\@nil}
\def\mn@eprint@arXiv#1{\href {http://arxiv.org/abs/#1} {{\tt arXiv:#1}}}
\def\mn@eprint@dblp#1{\href {http://dblp.uni-trier.de/rec/bibtex/#1.xml}
  {dblp:#1}}
\def\mn@eprint@#1:#2:#3:#4\@nil{\def\@tempa {#1}\def\@tempb {#2}\def\@tempc
  {#3}\ifx \@tempc \@empty \let \@tempc \@tempb \let \@tempb \@tempa \fi \ifx
  \@tempb \@empty \def\@tempb {arXiv}\fi \@ifundefined
  {mn@eprint@\@tempb}{\@tempb:\@tempc}{\expandafter \expandafter \csname
  mn@eprint@\@tempb\endcsname \expandafter{\@tempc}}}

\bibitem[\protect\citeauthoryear{{Agarwal} \& {Milosavljevi{\'c}}}{{Agarwal} \&
  {Milosavljevi{\'c}}}{2011}]{agarwal11}
{Agarwal} M.,  {Milosavljevi{\'c}} M.,  2011, \mn@doi [\apj]
  {10.1088/0004-637X/729/1/35}, \href
  {https://ui.adsabs.harvard.edu/abs/2011ApJ...729...35A} {729, 35}

\bibitem[\protect\citeauthoryear{{Alfaro-Cuello} et~al.,}{{Alfaro-Cuello}
  et~al.}{2019}]{alfaro19}
{Alfaro-Cuello} M.,  et~al., 2019, \mn@doi [\apj] {10.3847/1538-4357/ab1b2c},
  \href {https://ui.adsabs.harvard.edu/abs/2019ApJ...886...57A} {886, 57}

\bibitem[\protect\citeauthoryear{{Antonini}, {Barausse}  \& {Silk}}{{Antonini}
  et~al.}{2015}]{antonini15}
{Antonini} F.,  {Barausse} E.,   {Silk} J.,  2015, \mn@doi [\apj]
  {10.1088/0004-637X/812/1/72}, \href
  {https://ui.adsabs.harvard.edu/abs/2015ApJ...812...72A} {812, 72}

\bibitem[\protect\citeauthoryear{{Arca-Sedda} \&
  {Capuzzo-Dolcetta}}{{Arca-Sedda} \&
  {Capuzzo-Dolcetta}}{2014}]{arcaseddacapuzzo14}
{Arca-Sedda} M.,  {Capuzzo-Dolcetta} R.,  2014, \mn@doi [\mnras]
  {10.1093/mnras/stu1683}, \href
  {https://ui.adsabs.harvard.edu/abs/2014MNRAS.444.3738A} {444, 3738}

\bibitem[\protect\citeauthoryear{{Bekki}}{{Bekki}}{2007}]{bekki07}
{Bekki} K.,  2007, \mn@doi [\pasa] {10.1071/AS07008}, \href
  {https://ui.adsabs.harvard.edu/abs/2007PASA...24...77B} {24, 77}

\bibitem[\protect\citeauthoryear{{Bekki} \& {Freeman}}{{Bekki} \&
  {Freeman}}{2003}]{bekkifreeman03}
{Bekki} K.,  {Freeman} K.~C.,  2003, \mn@doi [\mnras]
  {10.1046/j.1365-2966.2003.07275.x}, \href
  {https://ui.adsabs.harvard.edu/abs/2003MNRAS.346L..11B} {346, L11}

\bibitem[\protect\citeauthoryear{{Bellazzini}, {Fusi Pecci}, {Montegriffo},
  {Messineo}, {Monaco}  \& {Rood}}{{Bellazzini} et~al.}{2002}]{bellazzini02}
{Bellazzini} M.,  {Fusi Pecci} F.,  {Montegriffo} P.,  {Messineo} M.,  {Monaco}
  L.,   {Rood} R.~T.,  2002, \mn@doi [\aj] {10.1086/340082}, \href
  {http://adsabs.harvard.edu/abs/2002AJ....123.2541B} {123, 2541}

\bibitem[\protect\citeauthoryear{{Bellazzini} et~al.,}{{Bellazzini}
  et~al.}{2008}]{bellazzini08}
{Bellazzini} M.,  et~al., 2008, \mn@doi [\aj] {10.1088/0004-6256/136/3/1147},
  \href {https://ui.adsabs.harvard.edu/abs/2008AJ....136.1147B} {136, 1147}

\bibitem[\protect\citeauthoryear{{B{\"o}ker}, {Laine}, {van der Marel},
  {Sarzi}, {Rix}, {Ho}  \& {Shields}}{{B{\"o}ker} et~al.}{2002}]{boker02}
{B{\"o}ker} T.,  {Laine} S.,  {van der Marel} R.~P.,  {Sarzi} M.,  {Rix} H.-W.,
   {Ho} L.~C.,   {Shields} J.~C.,  2002, \mn@doi [\aj] {10.1086/339025}, \href
  {https://ui.adsabs.harvard.edu/abs/2002AJ....123.1389B} {123, 1389}

\bibitem[\protect\citeauthoryear{{Buonanno}, {Corsi}, {Castellani}, {Marconi},
  {Fusi Pecci}  \& {Zinn}}{{Buonanno} et~al.}{1999}]{buonanno99}
{Buonanno} R.,  {Corsi} C.~E.,  {Castellani} M.,  {Marconi} G.,  {Fusi Pecci}
  F.,   {Zinn} R.,  1999, \mn@doi [\aj] {10.1086/301034}, \href
  {https://ui.adsabs.harvard.edu/abs/1999AJ....118.1671B} {118, 1671}

\bibitem[\protect\citeauthoryear{{Cannon}, {Croke}, {Bell}, {Hesser}  \&
  {Stathakis}}{{Cannon} et~al.}{1998}]{cannon98}
{Cannon} R.~D.,  {Croke} B.~F.~W.,  {Bell} R.~A.,  {Hesser} J.~E.,
  {Stathakis} R.~A.,  1998, \mn@doi [\mnras]
  {10.1046/j.1365-8711.1998.01671.x}, \href
  {https://ui.adsabs.harvard.edu/abs/1998MNRAS.298..601C} {298, 601}

\bibitem[\protect\citeauthoryear{{Capuzzo-Dolcetta} \&
  {Miocchi}}{{Capuzzo-Dolcetta} \& {Miocchi}}{2008}]{capuzzomiocchi08}
{Capuzzo-Dolcetta} R.,  {Miocchi} P.,  2008, \mn@doi [\mnras]
  {10.1111/j.1745-3933.2008.00501.x}, \href
  {https://ui.adsabs.harvard.edu/abs/2008MNRAS.388L..69C} {388, L69}

\bibitem[\protect\citeauthoryear{{Carretta} et~al.,}{{Carretta}
  et~al.}{2009a}]{carretta09a}
{Carretta} E.,  et~al., 2009a, \mn@doi [Astron. Astrophys.]
  {10.1051/0004-6361/200912096}, \href
  {http://adsabs.harvard.edu/abs/2009A%26A...505..117C} {505, 117}

\bibitem[\protect\citeauthoryear{{Carretta}, {Bragaglia}, {Gratton}  \&
  {Lucatello}}{{Carretta} et~al.}{2009b}]{carretta09b}
{Carretta} E.,  {Bragaglia} A.,  {Gratton} R.,   {Lucatello} S.,  2009b,
  \mn@doi [\aap] {10.1051/0004-6361/200912097}, \href
  {https://ui.adsabs.harvard.edu/abs/2009A&A...505..139C} {505, 139}

\bibitem[\protect\citeauthoryear{{Carretta} et~al.,}{{Carretta}
  et~al.}{2010}]{carretta10}
{Carretta} E.,  et~al., 2010, \mn@doi [\aap] {10.1051/0004-6361/201014924},
  \href {https://ui.adsabs.harvard.edu/abs/2010A&A...520A..95C} {520, A95}

\bibitem[\protect\citeauthoryear{{Chantereau}, {Salaris}, {Bastian}  \&
  {Martocchia}}{{Chantereau} et~al.}{2019}]{chantereau19}
{Chantereau} W.,  {Salaris} M.,  {Bastian} N.,   {Martocchia} S.,  2019,
  \mn@doi [\mnras] {10.1093/mnras/stz378}, \href
  {https://ui.adsabs.harvard.edu/abs/2019MNRAS.484.5236C} {484, 5236}

\bibitem[\protect\citeauthoryear{{D'Antona}, {Caloi}, {D'Ercole}, {Tailo},
  {Vesperini}, {Ventura}  \& {Di Criscienzo}}{{D'Antona}
  et~al.}{2013}]{dantona13}
{D'Antona} F.,  {Caloi} V.,  {D'Ercole} A.,  {Tailo} M.,  {Vesperini} E.,
  {Ventura} P.,   {Di Criscienzo} M.,  2013, \mn@doi [\mnras]
  {10.1093/mnras/stt1057}, \href
  {https://ui.adsabs.harvard.edu/abs/2013MNRAS.434.1138D} {434, 1138}

\bibitem[\protect\citeauthoryear{{Dalessandro}, {Salaris}, {Ferraro},
  {Cassisi}, {Lanzoni}, {Rood}, {Fusi Pecci}  \& {Sabbi}}{{Dalessandro}
  et~al.}{2011a}]{dalessandro11b}
{Dalessandro} E.,  {Salaris} M.,  {Ferraro} F.~R.,  {Cassisi} S.,  {Lanzoni}
  B.,  {Rood} R.~T.,  {Fusi Pecci} F.,   {Sabbi} E.,  2011a, \mn@doi [\mnras]
  {10.1111/j.1365-2966.2010.17479.x}, \href
  {https://ui.adsabs.harvard.edu/abs/2011MNRAS.410..694D} {410, 694}

\bibitem[\protect\citeauthoryear{{Dalessandro}, {Lanzoni}, {Beccari},
  {Sollima}, {Ferraro}  \& {Pasquato}}{{Dalessandro}
  et~al.}{2011b}]{dalessandro11}
{Dalessandro} E.,  {Lanzoni} B.,  {Beccari} G.,  {Sollima} A.,  {Ferraro}
  F.~R.,   {Pasquato} M.,  2011b, \mn@doi [\apj] {10.1088/0004-637X/743/1/11},
  \href {http://adsabs.harvard.edu/abs/2011ApJ...743...11D} {743, 11}

\bibitem[\protect\citeauthoryear{{Dalessandro}, {Salaris}, {Ferraro},
  {Mucciarelli}  \& {Cassisi}}{{Dalessandro} et~al.}{2013a}]{dalessandro13}
{Dalessandro} E.,  {Salaris} M.,  {Ferraro} F.~R.,  {Mucciarelli} A.,
  {Cassisi} S.,  2013a, \mn@doi [\mnras] {10.1093/mnras/sts644}, \href
  {https://ui.adsabs.harvard.edu/abs/2013MNRAS.430..459D} {430, 459}

\bibitem[\protect\citeauthoryear{{Dalessandro} et~al.,}{{Dalessandro}
  et~al.}{2013b}]{dalessandro13b}
{Dalessandro} E.,  et~al., 2013b, \mn@doi [\apj] {10.1088/0004-637X/778/2/135},
  \href {https://ui.adsabs.harvard.edu/abs/2013ApJ...778..135D} {778, 135}

\bibitem[\protect\citeauthoryear{{Dalessandro} et~al.,}{{Dalessandro}
  et~al.}{2014}]{dalessandro14}
{Dalessandro} E.,  et~al., 2014, \mn@doi [ApJl] {10.1088/2041-8205/791/1/L4},
  \href {http://adsabs.harvard.edu/abs/2014ApJ...791L...4D} {791, L4}

\bibitem[\protect\citeauthoryear{{Dalessandro}, {Ferraro}, {Massari},
  {Lanzoni}, {Miocchi}  \& {Beccari}}{{Dalessandro}
  et~al.}{2015}]{dalessandro15}
{Dalessandro} E.,  {Ferraro} F.~R.,  {Massari} D.,  {Lanzoni} B.,  {Miocchi}
  P.,   {Beccari} G.,  2015, \mn@doi [\apj] {10.1088/0004-637X/810/1/40}, \href
  {http://adsabs.harvard.edu/abs/2015ApJ...810...40D} {810, 40}

\bibitem[\protect\citeauthoryear{{Dalessandro}, {Lapenna}, {Mucciarelli},
  {Origlia}, {Ferraro}  \& {Lanzoni}}{{Dalessandro}
  et~al.}{2016}]{dalessandro16}
{Dalessandro} E.,  {Lapenna} E.,  {Mucciarelli} A.,  {Origlia} L.,  {Ferraro}
  F.~R.,   {Lanzoni} B.,  2016, \mn@doi [ApJ] {10.3847/0004-637X/829/2/77},
  \href {http://adsabs.harvard.edu/abs/2016ApJ...829...77D} {829, 77}

\bibitem[\protect\citeauthoryear{{Dalessandro} et~al.,}{{Dalessandro}
  et~al.}{2018}]{dalessandro18}
{Dalessandro} E.,  et~al., 2018, \mn@doi [\aap] {10.1051/0004-6361/201833650},
  \href {http://adsabs.harvard.edu/abs/2018A%26A...618A.131D} {618, A131}

\bibitem[\protect\citeauthoryear{{Dalessandro}, {Ferraro}, {Bastian},
  {Cadelano}, {Lanzoni}  \& {Raso}}{{Dalessandro} et~al.}{2019}]{dalessandro19}
{Dalessandro} E.,  {Ferraro} F.~R.,  {Bastian} N.,  {Cadelano} M.,  {Lanzoni}
  B.,   {Raso} S.,  2019, \mn@doi [\aap] {10.1051/0004-6361/201834011}, \href
  {https://ui.adsabs.harvard.edu/abs/2019A&A...621A..45D} {621, A45}

\bibitem[\protect\citeauthoryear{{Fahrion} et~al.,}{{Fahrion}
  et~al.}{2019}]{fahrion19}
{Fahrion} K.,  et~al., 2019, \mn@doi [\aap] {10.1051/0004-6361/201935832},
  \href {https://ui.adsabs.harvard.edu/abs/2019A&A...628A..92F} {628, A92}

\bibitem[\protect\citeauthoryear{{Feldmeier} et~al.,}{{Feldmeier}
  et~al.}{2014}]{feldmeier14}
{Feldmeier} A.,  et~al., 2014, \mn@doi [\aap] {10.1051/0004-6361/201423777},
  \href {https://ui.adsabs.harvard.edu/abs/2014A&A...570A...2F} {570, A2}

\bibitem[\protect\citeauthoryear{{Georgiev}, {Puzia}, {Goudfrooij}  \&
  {Hilker}}{{Georgiev} et~al.}{2010}]{georgiev10}
{Georgiev} I.~Y.,  {Puzia} T.~H.,  {Goudfrooij} P.,   {Hilker} M.,  2010,
  \mn@doi [\mnras] {10.1111/j.1365-2966.2010.16802.x}, \href
  {https://ui.adsabs.harvard.edu/abs/2010MNRAS.406.1967G} {406, 1967}

\bibitem[\protect\citeauthoryear{{Gilligan} et~al.,}{{Gilligan}
  et~al.}{2019}]{gilligan19}
{Gilligan} C.~K.,  et~al., 2019, arXiv e-prints, \href
  {http://adsabs.harvard.edu/abs/2019arXiv190401434G} {}

\bibitem[\protect\citeauthoryear{{Girardi} et~al.,}{{Girardi}
  et~al.}{2008}]{gira08}
{Girardi} L.,  et~al., 2008, \mn@doi [\pasp] {10.1086/588526}, \href
  {https://ui.adsabs.harvard.edu/abs/2008PASP..120..583G} {120, 583}

\bibitem[\protect\citeauthoryear{{Gratton}, {Carretta}  \&
  {Bragaglia}}{{Gratton} et~al.}{2012}]{gratton12}
{Gratton} R.~G.,  {Carretta} E.,   {Bragaglia} A.,  2012, \mn@doi [\aapr]
  {10.1007/s00159-012-0050-3}, \href
  {http://adsabs.harvard.edu/abs/2012A%26ARv..20...50G} {20, 50}

\bibitem[\protect\citeauthoryear{{Greco} et~al.,}{{Greco}
  et~al.}{2007}]{greco07}
{Greco} C.,  et~al., 2007, \mn@doi [\apj] {10.1086/522102}, \href
  {https://ui.adsabs.harvard.edu/abs/2007ApJ...670..332G} {670, 332}

\bibitem[\protect\citeauthoryear{{Guillard}, {Emsellem}  \&
  {Renaud}}{{Guillard} et~al.}{2016}]{guillard16}
{Guillard} N.,  {Emsellem} E.,   {Renaud} F.,  2016, \mn@doi [\mnras]
  {10.1093/mnras/stw1570}, \href
  {https://ui.adsabs.harvard.edu/abs/2016MNRAS.461.3620G} {461, 3620}

\bibitem[\protect\citeauthoryear{{Hardy}}{{Hardy}}{2002}]{hardy02}
{Hardy} E.,  2002, in {Geisler} D.~P.,  {Grebel} E.~K.,   {Minniti} D.,  eds,
  IAU Symposium Vol. 207, Extragalactic Star Clusters. p.~62

\bibitem[\protect\citeauthoryear{{Hartmann}, {Debattista}, {Seth}, {Cappellari}
   \& {Quinn}}{{Hartmann} et~al.}{2011}]{hartmann11}
{Hartmann} M.,  {Debattista} V.~P.,  {Seth} A.,  {Cappellari} M.,   {Quinn}
  T.~R.,  2011, \mn@doi [\mnras] {10.1111/j.1365-2966.2011.19659.x}, \href
  {https://ui.adsabs.harvard.edu/abs/2011MNRAS.418.2697H} {418, 2697}

\bibitem[\protect\citeauthoryear{{Hendricks}, {Koch}, {Walker}, {Johnson},
  {Pe{\~n}arrubia}  \& {Gilmore}}{{Hendricks} et~al.}{2014}]{hendricks14}
{Hendricks} B.,  {Koch} A.,  {Walker} M.,  {Johnson} C.~I.,  {Pe{\~n}arrubia}
  J.,   {Gilmore} G.,  2014, \mn@doi [\aap] {10.1051/0004-6361/201424645},
  \href {https://ui.adsabs.harvard.edu/abs/2014A&A...572A..82H} {572, A82}

\bibitem[\protect\citeauthoryear{{Hendricks}, {Boeche}, {Johnson}, {Frank},
  {Koch}, {Mateo}  \& {Bailey}}{{Hendricks} et~al.}{2016}]{hendricks16}
{Hendricks} B.,  {Boeche} C.,  {Johnson} C.~I.,  {Frank} M.~J.,  {Koch} A.,
  {Mateo} M.,   {Bailey} J.~I.,  2016, \mn@doi [\aap]
  {10.1051/0004-6361/201526996}, \href
  {https://ui.adsabs.harvard.edu/abs/2016A&A...585A..86H} {585, A86}

\bibitem[\protect\citeauthoryear{{Hollyhead} et~al.,}{{Hollyhead}
  et~al.}{2018}]{hollyhead18}
{Hollyhead} K.,  et~al., 2018, \mn@doi [\mnras] {10.1093/mnras/sty230}, \href
  {http://adsabs.harvard.edu/abs/2018MNRAS.476..114H} {476, 114}

\bibitem[\protect\citeauthoryear{{Hollyhead} et~al.,}{{Hollyhead}
  et~al.}{2019}]{hollyhead19}
{Hollyhead} K.,  et~al., 2019, \mn@doi [\mnras] {10.1093/mnras/stz317}, \href
  {http://adsabs.harvard.edu/abs/2019MNRAS.484.4718H} {484, 4718}

\bibitem[\protect\citeauthoryear{{Kacharov}, {Neumayer}, {Seth}, {Cappellari},
  {McDermid}, {Walcher}  \& {B{\"o}ker}}{{Kacharov} et~al.}{2018}]{kacharov18}
{Kacharov} N.,  {Neumayer} N.,  {Seth} A.~C.,  {Cappellari} M.,  {McDermid} R.,
   {Walcher} C.~J.,   {B{\"o}ker} T.,  2018, \mn@doi [\mnras]
  {10.1093/mnras/sty1985}, \href
  {https://ui.adsabs.harvard.edu/abs/2018MNRAS.480.1973K} {480, 1973}

\bibitem[\protect\citeauthoryear{{King}}{{King}}{1966}]{king66}
{King} I.~R.,  1966, \mn@doi [\aj] {10.1086/109857}, \href
  {https://ui.adsabs.harvard.edu/abs/1966AJ.....71...64K} {71, 64}

\bibitem[\protect\citeauthoryear{{Lamers}, {Kruijssen}, {Bastian}, {Rejkuba},
  {Hilker}  \& {Kissler-Patig}}{{Lamers} et~al.}{2017}]{lamers17}
{Lamers} H.~J.~G.~L.~M.,  {Kruijssen} J.~M.~D.,  {Bastian} N.,  {Rejkuba} M.,
  {Hilker} M.,   {Kissler-Patig} M.,  2017, \mn@doi [\aap]
  {10.1051/0004-6361/201731062}, \href
  {https://ui.adsabs.harvard.edu/abs/2017A&A...606A..85L} {606, A85}

\bibitem[\protect\citeauthoryear{{Larsen}, {Strader}  \& {Brodie}}{{Larsen}
  et~al.}{2012a}]{larsen12a}
{Larsen} S.~S.,  {Strader} J.,   {Brodie} J.~P.,  2012a, \mn@doi [\aap]
  {10.1051/0004-6361/201219897}, \href
  {https://ui.adsabs.harvard.edu/abs/2012A&A...544L..14L} {544, L14}

\bibitem[\protect\citeauthoryear{{Larsen}, {Brodie}  \& {Strader}}{{Larsen}
  et~al.}{2012b}]{larsen12}
{Larsen} S.~S.,  {Brodie} J.~P.,   {Strader} J.,  2012b, \mn@doi [\aap]
  {10.1051/0004-6361/201219895}, \href
  {https://ui.adsabs.harvard.edu/abs/2012A&A...546A..53L} {546, A53}

\bibitem[\protect\citeauthoryear{{Larsen}, {Brodie}, {Grundahl}  \&
  {Strader}}{{Larsen} et~al.}{2014}]{larsen14}
{Larsen} S.~S.,  {Brodie} J.~P.,  {Grundahl} F.,   {Strader} J.,  2014, \mn@doi
  [\apj] {10.1088/0004-637X/797/1/15}, \href
  {https://ui.adsabs.harvard.edu/abs/2014ApJ...797...15L} {797, 15}

\bibitem[\protect\citeauthoryear{{Letarte}, {Hill}, {Jablonka}, {Tolstoy},
  {Fran{\c{c}}ois}  \& {Meylan}}{{Letarte} et~al.}{2006}]{letarte06}
{Letarte} B.,  {Hill} V.,  {Jablonka} P.,  {Tolstoy} E.,  {Fran{\c{c}}ois} P.,
   {Meylan} G.,  2006, \mn@doi [\aap] {10.1051/0004-6361:20054439}, \href
  {https://ui.adsabs.harvard.edu/abs/2006A&A...453..547L} {453, 547}

\bibitem[\protect\citeauthoryear{{Lyubenova} et~al.,}{{Lyubenova}
  et~al.}{2013}]{lyubenova13}
{Lyubenova} M.,  et~al., 2013, \mn@doi [\mnras] {10.1093/mnras/stt414}, \href
  {https://ui.adsabs.harvard.edu/abs/2013MNRAS.431.3364L} {431, 3364}

\bibitem[\protect\citeauthoryear{{Mackey} \& {Gilmore}}{{Mackey} \&
  {Gilmore}}{2003}]{mackeygilmore03}
{Mackey} A.~D.,  {Gilmore} G.~F.,  2003, \mn@doi [\mnras]
  {10.1046/j.1365-8711.2003.06275.x}, \href
  {https://ui.adsabs.harvard.edu/abs/2003MNRAS.340..175M} {340, 175}

\bibitem[\protect\citeauthoryear{{Marino}, {Villanova}, {Piotto}, {Milone},
  {Momany}, {Bedin}  \& {Medling}}{{Marino} et~al.}{2008}]{marino08}
{Marino} A.~F.,  {Villanova} S.,  {Piotto} G.,  {Milone} A.~P.,  {Momany} Y.,
  {Bedin} L.~R.,   {Medling} A.~M.,  2008, \mn@doi [\aap]
  {10.1051/0004-6361:200810389}, \href
  {https://ui.adsabs.harvard.edu/abs/2008A&A...490..625M} {490, 625}

\bibitem[\protect\citeauthoryear{{Martocchia} et~al.,}{{Martocchia}
  et~al.}{2017}]{martocchia17}
{Martocchia} S.,  et~al., 2017, \mn@doi [MNRAS] {10.1093/mnras/stx660}, \href
  {http://adsabs.harvard.edu/abs/2017MNRAS.468.3150M} {468, 3150}

\bibitem[\protect\citeauthoryear{{Martocchia} et~al.,}{{Martocchia}
  et~al.}{2018}]{martocchia18a}
{Martocchia} S.,  et~al., 2018, \mn@doi [\mnras] {10.1093/mnras/stx2556}, \href
  {http://adsabs.harvard.edu/abs/2018MNRAS.473.2688M} {473, 2688}

\bibitem[\protect\citeauthoryear{{Martocchia} et~al.,}{{Martocchia}
  et~al.}{2019}]{martocchia19}
{Martocchia} S.,  et~al., 2019, \mn@doi [\mnras] {10.1093/mnras/stz1596}, \href
  {https://ui.adsabs.harvard.edu/abs/2019MNRAS.tmp.1528M} {p.~1528}

\bibitem[\protect\citeauthoryear{{McConnachie}}{{McConnachie}}{2012}]{mcconnachie12}
{McConnachie} A.~W.,  2012, \mn@doi [\aj] {10.1088/0004-6256/144/1/4}, \href
  {https://ui.adsabs.harvard.edu/abs/2012AJ....144....4M} {144, 4}

\bibitem[\protect\citeauthoryear{{Milone} et~al.,}{{Milone}
  et~al.}{2012}]{milone12}
{Milone} A.~P.,  et~al., 2012, \mn@doi [\aap] {10.1051/0004-6361/201016384},
  \href {http://adsabs.harvard.edu/abs/2012A%26A...540A..16M} {540, A16}

\bibitem[\protect\citeauthoryear{{Milone} et~al.,}{{Milone}
  et~al.}{2015}]{milone15}
{Milone} A.~P.,  et~al., 2015, \mn@doi [\mnras] {10.1093/mnras/stv829}, \href
  {https://ui.adsabs.harvard.edu/abs/2015MNRAS.450.3750M} {450, 3750}

\bibitem[\protect\citeauthoryear{{Milone} et~al.,}{{Milone}
  et~al.}{2017}]{milone17}
{Milone} A.~P.,  et~al., 2017, \mn@doi [\mnras] {10.1093/mnras/stw2531}, \href
  {http://adsabs.harvard.edu/abs/2017MNRAS.464.3636M} {464, 3636}

\bibitem[\protect\citeauthoryear{{Miocchi} et~al.,}{{Miocchi}
  et~al.}{2013}]{miocchi13}
{Miocchi} P.,  et~al., 2013, \mn@doi [\apj] {10.1088/0004-637X/774/2/151},
  \href {https://ui.adsabs.harvard.edu/abs/2013ApJ...774..151M} {774, 151}

\bibitem[\protect\citeauthoryear{{Mucciarelli}, {Origlia}, {Ferraro}  \&
  {Pancino}}{{Mucciarelli} et~al.}{2009}]{mucciarelli09}
{Mucciarelli} A.,  {Origlia} L.,  {Ferraro} F.~R.,   {Pancino} E.,  2009,
  \mn@doi [ApJ] {10.1088/0004-637X/695/2/L134}, \href
  {http://adsabs.harvard.edu/abs/2009ApJ...695L.134M} {695, L134}

\bibitem[\protect\citeauthoryear{{Nardiello} et~al.,}{{Nardiello}
  et~al.}{2018}]{nardiello18}
{Nardiello} D.,  et~al., 2018, \mn@doi [\mnras] {10.1093/mnras/sty2515}, \href
  {https://ui.adsabs.harvard.edu/abs/2018MNRAS.481.3382N} {481, 3382}

\bibitem[\protect\citeauthoryear{{Neumayer}, {Walcher}, {Andersen},
  {S{\'a}nchez}, {B{\"o}ker}  \& {Rix}}{{Neumayer} et~al.}{2011}]{neumayer11}
{Neumayer} N.,  {Walcher} C.~J.,  {Andersen} D.,  {S{\'a}nchez} S.~F.,
  {B{\"o}ker} T.,   {Rix} H.-W.,  2011, \mn@doi [\mnras]
  {10.1111/j.1365-2966.2011.18266.x}, \href
  {https://ui.adsabs.harvard.edu/abs/2011MNRAS.413.1875N} {413, 1875}

\bibitem[\protect\citeauthoryear{{Neumayer}, {Seth}  \& {Boeker}}{{Neumayer}
  et~al.}{2020}]{neumayer20}
{Neumayer} N.,  {Seth} A.,   {Boeker} T.,  2020, arXiv e-prints, \href
  {https://ui.adsabs.harvard.edu/abs/2020arXiv200103626N} {p. arXiv:2001.03626}

\bibitem[\protect\citeauthoryear{{Niederhofer} et~al.,}{{Niederhofer}
  et~al.}{2017a}]{niederhofer17a}
{Niederhofer} F.,  et~al., 2017a, \mn@doi [\mnras] {10.1093/mnras/stw2269},
  \href {https://ui.adsabs.harvard.edu/abs/2017MNRAS.464...94N} {464, 94}

\bibitem[\protect\citeauthoryear{{Niederhofer} et~al.,}{{Niederhofer}
  et~al.}{2017b}]{niederhofer17b}
{Niederhofer} F.,  et~al., 2017b, \mn@doi [MNRAS] {10.1093/mnras/stw3084},
  \href {http://adsabs.harvard.edu/abs/2017MNRAS.465.4159N} {465, 4159}

\bibitem[\protect\citeauthoryear{{Pietrinferni}, {Cassisi}, {Salaris}  \&
  {Castelli}}{{Pietrinferni} et~al.}{2004}]{pietrinferni04}
{Pietrinferni} A.,  {Cassisi} S.,  {Salaris} M.,   {Castelli} F.,  2004,
  \mn@doi [ApJ] {10.1086/422498}, \href
  {http://adsabs.harvard.edu/abs/2004ApJ...612..168P} {612, 168}

\bibitem[\protect\citeauthoryear{{Pietrinferni}, {Cassisi}, {Salaris}  \&
  {Castelli}}{{Pietrinferni} et~al.}{2006}]{pietrinferni06}
{Pietrinferni} A.,  {Cassisi} S.,  {Salaris} M.,   {Castelli} F.,  2006,
  \mn@doi [\apj] {10.1086/501344}, \href
  {https://ui.adsabs.harvard.edu/abs/2006ApJ...642..797P} {642, 797}

\bibitem[\protect\citeauthoryear{{Piotto} et~al.,}{{Piotto}
  et~al.}{2015}]{piotto15}
{Piotto} G.,  et~al., 2015, \mn@doi [\aj] {10.1088/0004-6256/149/3/91}, \href
  {https://ui.adsabs.harvard.edu/abs/2015AJ....149...91P} {149, 91}

\bibitem[\protect\citeauthoryear{{Rusakov}, {Monelli}, {Gallart}, {Fritz},
  {Ruiz-Lara}, {Bernard}  \& {Cassisi}}{{Rusakov} et~al.}{2020}]{rusakov20}
{Rusakov} V.,  {Monelli} M.,  {Gallart} C.,  {Fritz} T.~K.,  {Ruiz-Lara} T.,
  {Bernard} E.~J.,   {Cassisi} S.,  2020, arXiv e-prints, \href
  {https://ui.adsabs.harvard.edu/abs/2020arXiv200209714R} {p. arXiv:2002.09714}

\bibitem[\protect\citeauthoryear{{Saracino} et~al.,}{{Saracino}
  et~al.}{2018}]{saracino18}
{Saracino} S.,  et~al., 2018, \mn@doi [\apj] {10.3847/1538-4357/aac2c2}, \href
  {https://ui.adsabs.harvard.edu/abs/2018ApJ...860...95S} {860, 95}

\bibitem[\protect\citeauthoryear{{Schiavon}, {Caldwell}, {Conroy}, {Graves},
  {Strader}, {MacArthur}, {Courteau}  \& {Harding}}{{Schiavon}
  et~al.}{2013}]{schiavon13}
{Schiavon} R.~P.,  {Caldwell} N.,  {Conroy} C.,  {Graves} G.~J.,  {Strader} J.,
   {MacArthur} L.~A.,  {Courteau} S.,   {Harding} P.,  2013, \mn@doi [\apjl]
  {10.1088/2041-8205/776/1/L7}, \href
  {https://ui.adsabs.harvard.edu/abs/2013ApJ...776L...7S} {776, L7}

\bibitem[\protect\citeauthoryear{{Sills}, {Dalessandro}, {Cadelano},
  {Alfaro-Cuello}  \& {Kruijssen}}{{Sills} et~al.}{2019}]{sills19}
{Sills} A.,  {Dalessandro} E.,  {Cadelano} M.,  {Alfaro-Cuello} M.,
  {Kruijssen} J.~M.~D.,  2019, \mn@doi [\mnras] {10.1093/mnrasl/slz149}, \href
  {https://ui.adsabs.harvard.edu/abs/2019MNRAS.490L..67S} {490, L67}

\bibitem[\protect\citeauthoryear{{Stetson}}{{Stetson}}{1987}]{stetson87}
{Stetson} P.~B.,  1987, \mn@doi [PASP] {10.1086/131977}, \href
  {http://adsabs.harvard.edu/abs/1987PASP...99..191S} {99, 191}

\bibitem[\protect\citeauthoryear{{Strader}, {Brodie}, {Forbes}, {Beasley}  \&
  {Huchra}}{{Strader} et~al.}{2003}]{strader03}
{Strader} J.,  {Brodie} J.~P.,  {Forbes} D.~A.,  {Beasley} M.~A.,   {Huchra}
  J.~P.,  2003, \mn@doi [\aj] {10.1086/367599}, \href
  {https://ui.adsabs.harvard.edu/abs/2003AJ....125.1291S} {125, 1291}

\bibitem[\protect\citeauthoryear{{Tremaine}, {Ostriker}  \&
  {Spitzer}}{{Tremaine} et~al.}{1975}]{tremaine75}
{Tremaine} S.~D.,  {Ostriker} J.~P.,   {Spitzer} L. J.,  1975, \mn@doi [\apj]
  {10.1086/153422}, \href
  {https://ui.adsabs.harvard.edu/abs/1975ApJ...196..407T} {196, 407}

\bibitem[\protect\citeauthoryear{{Walcher} et~al.,}{{Walcher}
  et~al.}{2005}]{walcher05}
{Walcher} C.~J.,  et~al., 2005, \mn@doi [\apj] {10.1086/425977}, \href
  {https://ui.adsabs.harvard.edu/abs/2005ApJ...618..237W} {618, 237}

\bibitem[\protect\citeauthoryear{{Walcher}, {B{\"o}ker}, {Charlot}, {Ho},
  {Rix}, {Rossa}, {Shields}  \& {van der Marel}}{{Walcher}
  et~al.}{2006}]{walcher06}
{Walcher} C.~J.,  {B{\"o}ker} T.,  {Charlot} S.,  {Ho} L.~C.,  {Rix} H.~W.,
  {Rossa} J.,  {Shields} J.~C.,   {van der Marel} R.~P.,  2006, \mn@doi [\apj]
  {10.1086/505166}, \href
  {https://ui.adsabs.harvard.edu/abs/2006ApJ...649..692W} {649, 692}

\bibitem[\protect\citeauthoryear{{Wang} et~al.,}{{Wang} et~al.}{2019}]{wang19}
{Wang} M.~Y.,  et~al., 2019, \mn@doi [\apjl] {10.3847/2041-8213/ab14f5}, \href
  {https://ui.adsabs.harvard.edu/abs/2019ApJ...875L..13W} {875, L13}

\bibitem[\protect\citeauthoryear{{Webbink}}{{Webbink}}{1985}]{webbink85}
{Webbink} R.~F.,  1985, in {Goodman} J.,  {Hut} P.,  eds,  IAU Symposium Vol.
  113, Dynamics of Star Clusters. pp 541--577

\bibitem[\protect\citeauthoryear{{de Boer} \& {Fraser}}{{de Boer} \&
  {Fraser}}{2016}]{deboer16}
{de Boer} T.~J.~L.,  {Fraser} M.,  2016, \mn@doi [\aap]
  {10.1051/0004-6361/201527580}, \href
  {https://ui.adsabs.harvard.edu/abs/2016A&A...590A..35D} {590, A35}

\bibitem[\protect\citeauthoryear{{van Dokkum}}{{van
  Dokkum}}{2001}]{vandokkum01}
{van Dokkum} P.~G.,  2001, \mn@doi [PASP] {10.1086/323894}, \href
  {http://adsabs.harvard.edu/abs/2001PASP..113.1420V} {113, 1420}

\makeatother
\end{thebibliography}






\bsp	
\label{lastpage}

%

\end{document}